# Designing a Theorem Prover

Lawrence C Paulson
Computer Laboratory
University of Cambridge

May 1990

# Contents

# 1  Folderol: a simple theorem prover

Because many different forms of logic are applicable to Computer Science, a common question is — *How do I write a theorem prover?* This question can be answered with general advice. For instance, first do enough paper proofs to show that automation of your logic is both necessary and feasible. The question can also be answered with a survey of existing provers, as will be done in this chapter. But automatic theorem-proving involves a combination of theoretical and coding skills that is best illustrated by a case study. So this chapter presents a toy theorem prover, called *Folderol*, highlighting the key design issues. Code fragments are discussed in the text; a full program listing appears at the end of the chapter.

Folderol works in classical first-order logic. It follows an automatic strategy but is interactive. You can enter a conjecture, let Folderol work with it for a while, and finally modify or abandon the conjecture if it cannot be proved. There is no practical automatic proof procedure, even for most complete theories. Even if there were, interaction with the machine would still be necessary for exploring mathematics.

A theorem prover should offer some evidence that its proofs are valid. The Boyer and Moore [1979] theorem-prover prints an English summary of its reasoning, while Folderol prints a trace of the rules. Most LCF-style systems offer no evidence of correctness other than an obstinate insistence on playing by the rules [Paulson 1987]. If correctness is a matter of life and death, then a prover can be designed to output its proof for checking by a separate program.

Absolute correctness can never be obtained, even with a computer-checked proof. There are fundamental reasons for this. Any program may contain errors, including the one that checks the proofs; we have no absolutely reliable means by which to verify the verifier. In addition, a proof of correctness of a program relies on assumptions about the real world; it assumes that the computer works correctly, and usually assumes that the program is used properly. Hardware correctness proofs depend upon similar assumptions [Cohn 1989b]. Even in pure mathematics, the premises of a theorem are subject to revision [Lakatos 1976].

Although Folderol performs well on certain examples, its proof strategy suffers inherent limitations. The strategy would be even worse if we adopted modal or intuitionistic logic, and cannot easily be remedied. My hope is that, once you have studied this chapter, you will gain the confidence to implement more sophisticated strategies.

Folderol exploits many of the concepts of functional programming, although it is not pure. It is written in Standard ML, which is increasingly popular for theorem proving. Several recent books teach this language.

**Remark on quantifiers.**  The usual notation for quantifiers, seen elsewhere in this volume, gives the quantified variable the smallest possible scope. To save on parentheses, quantifiers in this chapter employ a dot convention, extending the scope of quantification as far as possible. So $\forall x.P \to Q$ abbreviates $\forall x.(P \to Q)$, and

$$\exists y.P \land \exists z.Q \lor R \text{ abbreviates } \exists y.(P \land (\exists z.(Q \lor R)))$$

## 1.1 Representation of rules

Given that Folderol will handle classical first-order logic, what formal system is best suited for automation? Gentzen's sequent calculus LK [Takeuti 1987] supports backwards proof in a natural way. A naïve process of working upwards from the desired conclusion yields many proofs.

Most rules act on one formula, and they come in pairs: one operating on the left and one on the right. We must inspect the rules carefully before choosing data structures. The quantifier rules, especially, constrain the representation of terms.

The classical sequent calculus deals with sequents of the form $\Gamma \vdash \Delta$. Here $\Gamma$ is the left part and $\Delta$ is the right part, and $\Gamma \vdash \Delta$ means 'if every formula in $\Gamma$ is true then some formula in $\Delta$ is true.' The sequent $A \vdash A$ (called a *basic* sequent) is trivially true. You may prefer to think in terms of refutation. Then $\Gamma \vdash \Delta$ is false if every formula in $\Gamma$ is true and every formula in $\Delta$ is false. If these conditions are contradictory then the sequent is true.

*Structural* rules, namely exchange, thinning, and contraction, make sequents behave like a consequence relation.

The *thinning* rules, in backwards proof, delete a formula from the goal. Since a formula should not be deleted unless it is known to be redundant, let us postpone all deletions. A goal is trivially true if it has the form $\Gamma \vdash \Delta$ where there exists a common formula $A \in \Gamma \cap \Delta$. Accepting these as basic sequents makes thinning rules unnecessary; the other formulae are simply thrown away.

The *exchange* rules swap two adjacent formulae, for traditionally $\Gamma$ and $\Delta$ are lists, not sets. Lists are also convenient in programming. We can do without exchange rules by ignoring the order of formulae in a sequent. In the rule

$$\frac{\Gamma \vdash \Delta, A \qquad \Gamma \vdash \Delta, B}{\Gamma \vdash \Delta, A \wedge B} \quad \wedge\text{:right}$$

there is no reason why $A \wedge B$ needs to be last. It can appear anywhere in the right part of the conclusion. For backwards proof: if the goal contains $A \wedge B$ anywhere in the right part then we can produce two subgoals, replacing $A \wedge B$ by $A$ and $B$ respectively. We can write the rule more concisely and clearly as follows, showing only formulae that change:

$$\frac{\vdash A \qquad \vdash B}{\vdash A \wedge B} \quad \wedge\text{:right}$$

The *contraction* rules, in backwards proof, duplicate a formula. Duplicating the $A \wedge B$, then applying the conjunction rule above, makes subgoals $\vdash A, A \wedge B$ and $\vdash B, A \wedge B$. These are equivalent to $\vdash A$ and $\vdash B$, so the $A \wedge B$ is redundant.

A case-by-case inspection of the rules reveals that the only formulae worth duplicating are $\forall x.A$ on the left and $\exists x.A$ on the right. Let us add contraction to the rules $\forall$:left and $\exists$:right. The rule $\forall$:left takes a goal containing $\forall x.A$ on the left and makes one subgoal by adding the formula $A[t/x]$ for some term $t$. The subgoal retains a copy of $\forall x.A$ so that the rule can be applied again for some other term. This example requires $n$ repeated uses of the quantified formula on terms $a$, $f(a)$,

...:
$$\forall x \,.\, P(x) \rightarrow P(f(x)) \;\vdash\; P(a) \rightarrow P(\underbrace{f(f(\cdots f}_{n \text{ times}}(a)\cdots)))$$

The rule ∃:right similarly retains the quantified formula in its subgoal.

Folderol uses these quantified formulae in rotation. It never throws them away. Thus even
$$\forall x.P(x) \vdash Q$$
makes Folderol run forever. If not for the re-use of quantified formulae, the search space would be finite. A theorem prover should only instantiate a quantifier when strictly necessary — but there is no effective test. First-order logic is undecidable.

## 1.2  Propositional logic

Propositional logic concerns the connectives ∧, ∨, →, ↔, and ¬. We take negation as primitive rather than defining ¬$A$ as $A \rightarrow \bot$. Although $A \leftrightarrow B$ means $(A \rightarrow B) \wedge (B \rightarrow A)$, performing this expansion could cause exponential blowup! The rules for ↔ permit natural reasoning.

Figure 1 shows the rules for each connective. In backwards proof each rule breaks down a non-atomic formula in the conclusion. If more than one rule is applicable, as in $A \wedge B \vdash B \wedge A$, which should be chosen? Some choices give shorter proofs than others. This proof of $A \wedge B \vdash B \wedge A$ uses ∧:left before ∧:right (working upwards):

$$\cfrac{\cfrac{A, B \vdash B \quad A, B \vdash A}{A, B \vdash B \wedge A}}{A \wedge B \vdash B \wedge A} \begin{array}{l} \wedge\text{:right} \\ \wedge\text{:left} \end{array}$$

If ∧:right is used first then ∧:left must be used twice. In larger examples the difference can be exponential.

$$\cfrac{\cfrac{A, B \vdash B}{A \wedge B \vdash B} \wedge\text{:left} \quad \cfrac{A, B \vdash A}{A \wedge B \vdash A} \wedge\text{:left}}{A \wedge B \vdash B \wedge A} \wedge\text{:right}$$

Rules can be chosen to minimize the proliferation of subgoals. If a goal is a basic sequent then trying other rules seems pointless.[1] Let us say that the *cost* of a rule is the number of premises it has. A rule with one premise is cheaper than a rule with two premises; ∧:left is cheaper than ∧:right.

The obvious representation of the goal $\Gamma \vdash \Delta$, a pair of lists, is awkward for choosing the cheapest rule. A Folderol goal is a list of triples ordered by cost. Each triple contains a formula, a 'side' (either Left or Right), and a cost.

---
[1] But once we introduce unification, we have to be careful. Solving a goal by unification may instantiate variables in other goals, possibly rendering them unprovable.

|   | *left* | *right* |
|---|---|---|
| $\wedge$ | $\dfrac{A, B \vdash}{A \wedge B \vdash}$ | $\dfrac{\vdash A \quad \vdash B}{\vdash A \wedge B}$ |
| $\vee$ | $\dfrac{A \vdash \quad B \vdash}{A \vee B \vdash}$ | $\dfrac{\vdash A, B}{\vdash A \vee B}$ |
| $\rightarrow$ | $\dfrac{\vdash A \quad B \vdash}{A \rightarrow B \vdash}$ | $\dfrac{A \vdash B}{\vdash A \rightarrow B}$ |
| $\leftrightarrow$ | $\dfrac{A, B \vdash \quad \vdash A, B}{A \leftrightarrow B \vdash}$ | $\dfrac{A \vdash B \quad B \vdash A}{\vdash A \leftrightarrow B}$ |
| $\neg$ | $\dfrac{\vdash A}{\neg A \vdash}$ | $\dfrac{A \vdash}{\vdash \neg A}$ |

Figure 1: Propositional logic

|   | *left* | *right* |
|---|---|---|
| $\forall$ | $\dfrac{\forall x.A, A[t/x] \vdash}{\forall x.A \vdash}$ | $\dfrac{\vdash A[a/x]}{\vdash \forall x.A}*$ |
| $\exists$ | $\dfrac{A[a/x] \vdash}{\exists x.A \vdash}*$ | $\dfrac{\vdash \exists x.A, A[t/x]}{\vdash \exists x.A}$ |

\* proviso: the parameter $a$ must not appear in the conclusion

Figure 2: Quantifier rules

## 1.3 Quantifiers and unification

First-order logic extends propositional logic with the quantifiers $\forall$ and $\exists$, as well as variables and other terms. Examination of the rules (Figure 2) reveals several difficulties involving terms.

In backwards proof, $\forall$:left and $\exists$:right the term $t$ is unspecified. It could be any combination of variables, constants, and functions. The possibilities are infinite. Why not defer the choice? At some future time, perhaps the correct choice will be obvious. Folderol uses a special kind of variable, really a meta-variable, to stand for an unspecified term. Such variables are written with a question mark: ?$a$, ?$b$, ?$c$, .... A backwards $\forall$:left or $\exists$:right inserts a fresh meta-variable ?$b$ in place of $t$ in the subgoal.

Whenever a new subgoal is produced, Folderol tries to change it to a basic sequent. It looks for a left-formula $A_1$ and right-formula $A_2$ such that replacing meta-variables by properly chosen terms turns both into the same formula $A$. We say $A$ is a *common instance* of $A_1$ and $A_2$.

The process of making such choices is called *unification*. For example, $R(?a, f(x))$ and $R(g(?b), ?b)$ both become $R(g(f(x)), f(x))$ replacing ?$a$ by $g(f(x))$ and ?$b$ by $f(x)$. This assignment is written ?$a \mapsto g(f(x))$, ?$b \mapsto f(x)$. Similarly, $R(?a, f(?c))$ and $R(g(?d), ?b)$ have $R(g(?d), f(?c))$ as a common instance. The resulting formula contains meta-variables ?$c$ and ?$d$ which can be replaced later.

No formula is a common instance of $P(?a)$ and $Q(f(y))$ if $P$ and $Q$ are distinct. More subtly, $P(?a)$ and $P(f(?a))$ have no common instance: no term $t$ is identical to $f(t)$ unless we admit the infinite term $f(f(f(\cdots)))$. Cycles can arise indirectly, as in unifying $R(?a, f(?a))$ and $R(?b, ?b)$.

## 1.4 Parameters in quantifier rules

Substitution should never cause a free variable to become bound. This error, called *variable capture*, permits unsound inferences such as the following:

$$\frac{\vdash \forall x.\exists y.y \neq x}{\vdash \exists y.y \neq y} \quad \forall\text{:right}$$

While substituting $y$ for $x$ in the subformula $\exists y.y \neq x$, the free variable $y$ becomes bound by the quantifier $\exists y$. In any model with two distinct individuals, the premise of the inference is true while the conclusion is clearly false. Renaming the bound variable $y$ to $z$ yields a correct conclusion, $\vdash \exists z.z \neq y$.

Variable capture can be prevented by renaming bound variables during substitution, but the algorithms to do this are complicated and liable to obscure errors. It is simpler to abandon the idea that a free variable is a variable that happens not to be bound. Instead have disjoint sets of parameters $a$, $b$, $c$, ..., and variables $x$, $y$, $z$, ..., where parameters may not be bound and variables must be bound.

Many logicians maintain this distinction between bound variables and parameters [Prawitz 1965]. Observe that a subformula may, under this distinction, not be a legal formula by itself. For instance, $\forall x.\exists y.y \neq x$ has the subformula $\exists y.y \neq x$, where $x$ appears unbound. We must replace the bound variable $x$ by a parameter, say $a$. Applying $\forall$:right to $\vdash \forall x.\exists y.y \neq x$ produces the conclusion $\vdash \exists y.y \neq a$.

### 1.4.1 Enforcement of provisos of quantifier rules

The proviso of $\forall$:right and $\exists$:left — that $a$ must not appear in the conclusion — ensures that $a$ can denote an arbitrary value. It may seem that if we choose a fresh parameter every time we use these rules, there is no danger of its appearing in the conclusion. But instantiation of meta-variables can change the conclusion, adding new parameters.

For example, $\forall x.R(x,x) \vdash \exists y.\forall x.R(x,y)$ is not valid. (To see this, let $R(x,y)$ mean $x = y$.) A derivation using meta-variables might be (deleting some quantified formulae)

$$\frac{\dfrac{\dfrac{\dfrac{R(?c,?c) \vdash R(b,?a)}{\forall x.R(x,x) \vdash R(b,?a)} \;\forall\text{:left}}{\forall x.R(x,x) \vdash \forall x.R(x,?a)} \;\forall\text{:right}}{\forall x.R(x,x) \vdash \exists y.\forall x.R(x,y)} \;\exists\text{:right}}{}$$

It appears that replacing both $?c$ and $?a$ by $b$ completes the proof, with top sequent $R(b,b) \vdash R(b,b)$. But then the second inference is nonsense:

$$\frac{\cdots \vdash R(b,b)}{\cdots \vdash \forall x.R(x,b)} \;\forall\text{:right}$$

Assigning $b$ to $?a$ sneaks $b$ into the conclusion of the rule $\forall$:right, violating its proviso. Let us say '$b$ depends on $?a$', which means $b$ must differ from the parameters in any term substituted for $?a$.

To apply $\forall$:right and $\exists$:left, choose a fresh parameter $b$ that depends on all the meta-variables in the goal, say $?a_1, \ldots, ?a_n$. To emphasize the dependence we write the parameter as $b[?a_1, \ldots, ?a_n]$. Compare with the function application $f(?a_1, \ldots, ?a_n)$ — parameters resemble Skolem terms, but do not use fictitious functions. Logically, $b[?a_1, \ldots, ?a_n]$ is like a Henkin constant (Barwise [1977], page 30).

Variable $?a$ is unifiable with term $t$ just if $?a$ does not occur in $t$. This *occurs check* can be extended to prevent invalid assignments to meta-variables. Folderol associates with every parameter the list of meta-variables it depends on: in the derivation above, the parameter is $b[?a]$. Assigning $b[?a]$ to $?a$ creates an obvious circularity. If $t$ contains $b[?a]$ then the occurs check will find that it depends on $?a$. The assignment $?a \mapsto t$, which would make $b$ appear in the conclusion, is forbidden.

For an exercise, verify that other possible proofs of $\forall x.R(x,x) \vdash \exists y.\forall x.R(x,y)$ fail.

Compare with a derivation of the valid sequent $\forall x.R(x,x) \vdash \forall x.\exists y.R(x,y)$:

$$\frac{\dfrac{\dfrac{\dfrac{R(?c,?c) \vdash R(a,?b)}{\forall x.R(x,x) \vdash R(a,?b)} \;\forall\text{:left}}{\forall x.R(x,x) \vdash \exists y.R(a,y)} \;\exists\text{:right}}{\forall x.R(x,x) \vdash \forall x.\exists y.R(x,y)} \;\forall\text{:right}}{}$$

Here the parameter $a$ does not depend on meta-variables $?b$ and $?c$ since these appear above it. Assigning $?c \mapsto a$ and $?b \mapsto a$ produces a valid proof:

$$\cfrac{\cfrac{\cfrac{R(a,a) \vdash R(a,a)}{\forall x.R(x,x) \vdash R(a,a)}\ \forall\text{:left}}{\forall x.R(x,x) \vdash \exists y.R(a,y)}\ \exists\text{:right}}{\forall x.R(x,x) \vdash \forall x.\,\exists y.R(x,y)}\ \forall\text{:right}$$

### 1.4.2 No second-order dependence

So each new parameter depends on all the meta-variables in the goal. But if the goal contains some other parameter $b[?a]$ this does not mean it contains $?a$. Consider this proof:

$$\cfrac{\vdash P(?a,b[?a]) \qquad \cfrac{\cfrac{\vdash Q(b[?a],c)}{\vdash \forall y.Q(b[?a],y)}\ \forall\text{:right}}{\vdash P(?a,b[?a]) \land (\forall y.Q(b[?a],y))}\ \land\text{:right}}{\vdash \forall x.\,P(?a,x) \land (\forall y.Q(x,y))}\ \forall\text{:right}$$

Here $b$ depends on $?a$, sequent $\vdash \forall y.Q(b[?a],y)$ contains $b[?a]$ but not $?a$, and $c$ does not depend on $?a$. The assignment $?a \mapsto c$ is unlikely to occur, for it takes $c$ outside of its natural scope. But this assignment would not violate the proviso of any rule. Only variables actually present in the conclusion affect the proviso. Thus Folderol does not make $c$ depend on $?a$.

Stricter conditions on the use of meta-variables can keep parameters within their scope. Folderol generates distinct meta-variables and only assigns $?a \mapsto t$ when all the symbols in $?a$ and $t$ are present in a single goal.

### 1.4.3 Summary

Let us restate the procedure for enforcing provisos of the rules $\forall$:right and $\exists$:left.

- When applying either rule, choose a fresh parameter — one that appears nowhere else in the proof — and make it depend on all meta-variables present in the conclusion.

- Allow the assignment $?a \mapsto t$ only if $t$ does not depend on $?a$. This ensures that each parameter $b$ in $t$ does not depend on $?a$, so the assignment respects the proviso of the rule that created $b$.

- To perform the assignment $?a \mapsto t$, replace $?a$ by $t$ throughout the proof. Replace each parameter depending on $?a$ by one depending on the meta-variables in $t$.

A lemma about LK justifies the substitution of terms for meta-variables under suitable conditions [Takeuti 1987, Lemma 2.11]. There is no logical need to distinguish meta-variables from parameters. But the distinction helps us — and Folderol — remember which variables are candidates for substitution.

# 2 Basic data structures and operations

Only now, having studied proof construction, can we choose a representation of terms and formulae. We will start to look at ML code including type definitions and substitution functions.

## 2.1 Terms

The function application $f(t_1, \ldots, t_n)$, where $f$ is an $n$-place function symbol and the $t_i$ are terms, is represented by the name of the function paired with the list of its arguments. There are $n$-place function symbols for all $n \geq 0$, where 0-place function symbols are constant symbols.

The three kinds of variables — bound variables, parameters, and meta-variables — require a system of names. In many-sorted logic, variables would also have a sort.

Meta-variables (henceforth called just 'variables') have a name, a character string. Names could be more complex: an integer subscript would allow quick renaming of variables.

Each parameter has a name and is paired with a list of variables (possibly empty) representing provisos of quantifier rules. So a parameter has the form $b[?a_1, \ldots, ?a_n]$. Alternatively we could maintain a table to pair each parameter $b$ with a list $?a_1, \ldots, ?a_n$.

Bound variables are distinguished from parameters using an approach [de Bruijn 1972] that also identifies expressions that are equivalent up to bound variable renaming. For example, $\forall x.\forall y.P(x)$ is equivalent to $\forall y.\forall x.P(y)$ but differs from $\forall x.\forall y.P(y)$, $\forall x.\forall x.P(x)$, and $\forall x.\forall y.P(a)$.

De Bruijn eliminates all bound variable names. The binding occurrence (after a quantifier in first-order logic) of a variable is dropped. Every other occurrence is represented by a non-negative integer, namely the number of enclosing quantifiers between the variable occurrence and its quantifier. These numbers are called de Bruijn indices and are written $\underline{0}$, $\underline{1}$, etc. Different occurrences of a variable may have different indices, and vice versa. Examples:

| | | |
|---|---|---|
| $\forall x.\forall y.P(x)$ | $\forall\forall P(\underline{1})$ | two equivalent formulae |
| $\forall y.\forall x.P(y)$ | $\forall\forall P(\underline{1})$ | |
| $\forall x.\forall y.P(y)$ | $\forall\forall P(\underline{0})$ | two equivalent formulae |
| $\forall x.\forall x.P(x)$ | $\forall\forall P(\underline{0})$ | |
| $\forall x.\forall y.P(a)$ | $\forall\forall P(a)$ | |
| $\exists x.((\exists y.Q(y)) \wedge S(x))$ | $\exists((\exists Q(\underline{0})) \wedge S(\underline{0}))$ | different variables, same index |
| $\forall x.(Q(x) \vee \exists z.R(z,x))$ | $\forall(Q(\underline{0}) \vee \exists R(\underline{0},\underline{1}))$ | same variables, different indices |

De Bruijn developed this representation for the $\lambda$-calculus, where $\lambda$-notation can bind variables in terms. A $\beta$-reduction may move a subterm into the scope of another $\lambda$. For example, $\lambda z.(\lambda x.\lambda y.x)(az)$ goes by $\beta$-conversion to $\lambda z.(\lambda y.az)$. The variable $z$ is free in $(az)$, which is substituted for $x$ in $\lambda y.x$. We must adjust the index in $(a\underline{0})$ as it goes inside the $\lambda y$:

$$\lambda z.(\lambda x.\lambda y.x)(az) \qquad \lambda(\lambda\lambda\underline{1})(a\underline{0})$$
$$\lambda z.(\lambda y.az) \qquad \lambda(\lambda a\underline{1})$$

De Bruijn [1972] gives simple algorithms for substitution, etc. They are especially simple for first-order logic, where terms cannot bind variables.

## 2.2 Formulae

There are three kinds of formula.

An atomic formula $P(t_1, \ldots, t_n)$, where $P$ is an $n$-place predicate symbol and the $t_i$ are terms, is represented by the name of the predicate and the list of its arguments. There are $n$-place predicate symbols for all $n \geq 0$.

A formula $A \wedge B$, $A \vee B$, $A \rightarrow B$, $A \leftrightarrow B$, or $\neg A$, where $A$ and $B$ are formulae, is represented by the connective and the list of subformulae: $[A, B]$ or $[A]$. Other connectives, like modal operators $\square$ and $\diamond$, can be represented similarly. This gives easy access to the subformulae. Most formula operations do not have a separate case for each connective.

A quantified formula $\forall x.A$ or $\exists x.A$, where the body $A$ is a formula and $x$ is a bound variable, is represented by the type of quantifier and the body. De Bruijn's notation does not require the name of the bound variable, but Folderol keeps it for printing results.

The discussion leads to the ML type definitions:

```
datatype term =
    Var   of string
  | Param of string * string list
  | Bound of int
  | Fun   of string * term list;

datatype form =
    Pred  of string * term list
  | Conn  of string * form list
  | Quant of string * string * form;
```

## 2.3 Abstraction and substitution

Two operations involving bound variables in formulae are abstraction and substitution.

- Abstraction maps $A$ to $\forall x.A[x/t]$ (or $\exists x.A[x/t]$), replacing all occurrences of the term $t$ by the bound variable $x$.

- Substitution maps $\forall x.A$ (or $\exists x.A$) to $A[t/x]$, replacing all occurrences of the bound variable $x$ by the term $t$.

Why is abstraction important? Recall that free variables are forbidden under de Bruijn notation. In $\forall x.A$ and $\exists x.A$, if $x$ is free in $A$ then the formula $A$ standing alone is not well-formed. So we cannot make a quantified formula by attaching $\forall x$ to the illegal formula $A$. Instead we use abstraction: if $A$ is a formula, $t$ is a term, and $x$ is a bound variable not present in $A$, then $\forall x.A[x/t]$ and $\exists x.A[x/t]$ are formulae.

Both abstraction and substitution start with bound variable index $\underline{0}$. This is incremented whenever a quantifier is encountered.

Abstraction replaces occurrences of $t$ by a bound variable. It calls the operation $\mathrm{abs}(i)$ for index $i \geq 0$, and $\mathrm{abs}(i)$ of a quantified formula involves $\mathrm{abs}(i+1)$ of the body. Clearly $\mathrm{abs}(i)$ of a conjunction, disjunction, ... involves $\mathrm{abs}(i)$ of each subformula. For predicates, $\mathrm{abs}(i)$ involves replacing $t$ by $\underline{i}$ in the arguments.

Example: abstract $(\forall z.R(z,a)) \leftrightarrow P(a)$ over $a$ to get $\exists x.(\forall z.R(z,x)) \leftrightarrow P(x)$:

$$\begin{aligned}
\mathrm{abs}(0)[(\forall R(\underline{0},a)) \leftrightarrow P(a)] &= \mathrm{abs}(0)[\forall R(\underline{0},a)] \leftrightarrow \mathrm{abs}(0)[P(a)] \\
&= (\forall \mathrm{abs}(1)[R(\underline{0},a)] \leftrightarrow P(\underline{0})) \\
&= (\forall R(\underline{0},\underline{1})) \leftrightarrow P(\underline{0})
\end{aligned}$$

Having computed $\mathrm{abs}(0)$ we attach the quantifier, getting the de Bruijn representation of the result:

$$\exists((\forall R(\underline{0},\underline{1})) \leftrightarrow P(\underline{0}))$$

Substitution replaces occurrences of the bound variable by $t$, having detached the outer quantifier. It calls the operation $\mathrm{subst}(i)$ for index $i \geq 0$. As above, $\mathrm{subst}(i)$ of a quantifier involves $\mathrm{subst}(i+1)$, while connectives do not require incrementing $i$ and predicates require replacing $\underline{i}$ by $t$ in the arguments. Substituting $a$ for $x$ in the body of $\exists x.(\forall z.R(z,x)) \leftrightarrow P(x)$ produces a computation almost identical to the previous example.

For the arguments of predicates both abstraction and substitution replace one term by another. This suggests a separate replacement operation on terms. This need not adjust indices since there is no $\lambda$-binding. The code for replacement, abstraction, and substitution is simple:

```
fun replace_term (u,new) t =
    if t=u then new else
    case t of Fun(a,ts) => Fun(a, map (replace_term(u,new)) ts)
            | _ => t;

fun abstract t =
    let fun abs i (Pred(a,ts)) =
                  Pred(a, map (replace_term (t, Bound i)) ts)
          | abs i (Conn(b,As)) = Conn(b, map (abs i) As)
          | abs i (Quant(q,b,A)) = Quant(q, b, abs (i+1) A)
    in  abs 0  end;

fun subst_bound t =
    let fun subst i (Pred(a,ts)) =
                  Pred(a, map (replace_term (Bound i, t)) ts)
          | subst i (Conn(b,As)) = Conn(b, map (subst i) As)
          | subst i (Quant(q,b,A)) = Quant(q, b, subst (i+1) A)
    in  subst 0  end;
```

Observe the use of the functional `map` to apply a function over a list, getting a list of results. This is used to handle the arguments of functions and predicates and the subformulae of connectives.

## 2.4 Parsing and printing

A good quarter of Folderol is concerned with parsing and printing of formulae. Let us pass over this quickly.

Abstraction is used only in the parser: given an input string for $\forall x.A$ it parses $A$ treating $x$ as a constant, then abstracts over $x$. This is hidden in the function

```
fun makeQuant q b A = Quant(q, b, abstract (Fun(b,[])) A);
```

Recall that a constant is a 0-place function.

Similarly the printer, given the formula $\forall x.A$, substitutes a constant named $x$ for the bound variable in $A$. The output is misleading if the body already contains a constant $x$. A better printer would make sure the name was unique.

Identifiers are sequences of letters or digits. Examples of terms include

| | |
|---|---|
| `r 12 banana` | constants |
| `?s ?12 ?apple` | variables |
| `f(x,?y) succ(succ(0))` | function applications |

The parser is crude. It accepts `f(f(f),f(f,f),f)`, for there is no table of function names. Also parameters cannot be expressed: an identifier by itself is taken for a constant.[2] There is no notation for bound variables, but none is needed.

Logical symbols are written as follows using computer characters:

|       |       |
|-------|-------|
| ~     | ¬     |
| &     | ∧     |
| \|    | ∨     |
| --> <-> | → ↔ |
| ALL EXISTS | ∀ ∃ |
| \|-   | ⊢     |

They appear in order of decreasing precedence; if precedence is equal, infixes associate to the right. Because quantifiers incorporate the dot notation, their scope extends as far as possible to the right. Enclose a quantified formula in parentheses if it is an operand of a connective.

For example, these theorems are provable by Folderol:

$$(P \lor (Q \land R)) \leftrightarrow ((P \lor Q) \land (P \lor R))$$

$$((P \to Q) \to P) \to P$$

$$\neg(\exists x.\forall y.F(x,y) \leftrightarrow \neg F(y,y))$$

$$\exists xy.P(x,y) \to \forall xy.P(x,y)$$

$$\exists x.\forall yz.(P(y) \to Q(z)) \to (P(x) \to Q(x))$$

Here they are in computer syntax:

```
P | (Q & R)   <-> (P | Q) & (P | R)
((P-->Q) --> P)   -->  P
~ (EXISTS x. ALL y. F(x,y) <-> ~ F(y,y))
EXISTS x. EXISTS y. P (x,y) -->   (ALL x. ALL y. P(x,y))
EXISTS x. ALL y. ALL z. (P(y)-->Q(z)) --> (P(x)-->Q(x))
```

# 3 Unification

The unification of terms $t$ and $u$ amounts to solving the equation $t = u$ over the (meta) variables in the terms. In the standard situation, all parameters and functions are distinct — the only equalities that hold are reflexivity ($t = t$) and the substitution of equals for equals. Thus $f(?a) = b$, $a = b$, and $f(?c) = g(?a, ?b)$ are each unsolvable, while each solution of $g(t_1, t_2) = g(u_1, u_2)$ solves $t_1 = u_1$ and $t_2 = u_2$ simultaneously.

In general we must solve a set of equations $\{t_1 = u_1, \ldots, t_n = u_n\}$. Equations are considered one at a time. Given an equation, there are nine cases: each term may be a function application, a parameter, or a variable. (Standard unification does not permit bound variables.) Most cases are unsolvable. If one term is a parameter then the other term must be the same parameter. If one term is a function application then the other must be an application of the same function, with the same number of arguments. The equation

$$f(t'_1, \ldots, t'_m) = f(u'_1, \ldots, u'_m)$$

is replaced by the equations

$$t'_1 = u'_1 \quad \ldots \quad t'_m = u'_m$$

---

[2]This limitation happily keeps the original goal free of parameters; these could otherwise clash with parameters created for quantifier rules.

The crucial case is an equation between a variable and a term: $?a = t$ or $t = ?a$. If $t$ is identical to $?a$ then we can drop the equation. Otherwise we must perform the *occurs check*. If $?a$ occurs in $t$ then the equation has no solution, for no term can properly contain itself. If $?a$ does not occur in $t$ then setting $?a \mapsto t$ solves the equation. In $?a = ?b$ either variable may be set to the other.

An equation is solved when it has been simplified to the form variable=term. Each solution must be substituted in the other equations, since we are solving them simultaneously. Then another equation is chosen. The process fails if an unsolvable equation is found and finishes when the entire set is solved. A solution set (or *unifier*) has the form $\{?a_1 = u_1, \ldots, ?a_k = u_k\}$, where the $?a_i$ are distinct and appear in none of the $u_j$. A unifier amounts to a substitution for $\{?a, \ldots, ?a_k\}$.

## 3.1 Examples

Here is an unsolvable example. The solutions of $g(a, c) = g(?b, ?b)$ are precisely solutions of the set of equations

$$a = ?b \quad\quad c = ?b$$

Substitution for $?b$ gives

$$?b = a \quad\quad c = a$$

and the second equation is unsolvable (two distinct constants).

Similarly $g(?a, f(?a)) = g(?b, ?b)$ reduces to

$$?a = ?b \quad\quad f(?a) = ?b$$

Substitution for $?a$ in the second equation makes it unsolvable (occurs check):

$$?a = ?b \quad\quad f(?b) = ?b$$

The equation $h(?a, f(?a), ?d) = h(g(0, ?d), ?b, ?c)$ reduces to

$$?a = g(0, ?d) \quad\quad f(?a) = ?b \quad\quad ?d = ?c$$

The first equation is solved; substituting it elsewhere gives

$$?a = g(0, ?d) \quad\quad f(g(0, ?d)) = ?b \quad\quad ?d = ?c$$

The second equation is solved:

$$?a = g(0, ?d) \qquad ?b = f(g(0, ?d)) \qquad ?d = ?c$$

Success! Solving the third equation gives the unifier

$$?a = g(0, ?c) \qquad ?b = f(g(0, ?c)) \qquad ?d = ?c$$

Infinitely many unifiers are instances of this, replacing $?c$ by any term.

Reversing the third equation to $?c = ?d$ above gives an equivalent unifier:

$$?a = g(0, ?d) \qquad ?b = f(g(0, ?d)) \qquad ?c = ?d$$

Since $?c = ?d$ exchanging the variables has no significance.

The method sketched above finds the *most general* unifier, one that includes all other unifiers as special cases. The equations may attempted in any order without significantly affecting the outcome. If a unifier exists then a most general unifier will be found. Two most general unifiers can differ up to variable renaming, as with $?c$ and $?d$ above.

## 3.2 Parameter dependencies in unification

Recall that a parameter $b$ generated in a quantifier rule has the form $b[?a_1, \ldots, ?a_n]$, where the $?a_i$ are all the variables in the conclusion. The occurs check (now a dependency check) finds that every $?a_i$ occurs in $b[?a_1, \ldots, ?a_n]$. In a parameter, assigning $?a_i \mapsto t$ replaces $?a_i$ by the variables in present in $t$.

For example, the set

$$?a = g(?c) \qquad ?c = b[?a]$$

reduces to the unsolvable

$$?a = g(?c) \qquad ?c = b[?c]$$

since $?c$ appears in $g(?c)$.

Contrast with this example, replacing $g(?c)$ by the parameter $d[?c]$. The set

$$?a = d[?c] \qquad ?c = b[?a]$$

is solvable since $?c$ does not actually appear in $d$ (thus $b[d[?c]]$ is just $b$):

$$?a = d[?c] \qquad ?c = b$$

This strange situation cannot happen in a real proof. If $?a$ and $?c$ are present then one, say $?a$, must have appeared first. Parameters $b[?a], d[?c]$ could never arise at once, instead perhaps $b[?c], d[?a, ?c]$ or $b[?a], d$.

The following set is also unifiable:

$$?a = d \qquad ?c = b[?a]$$

It corresponds to a proof where (going upwards) $d$, $?a$, $b[?a]$, and finally $?c$ appear.

## 3.3 The environment

It is inefficient to substitute throughout the equations every time a variable is assigned. Instead, the algorithm maintains an environment

$$[(?a_k, t_k), \ldots, (?a_1, t_1)]$$

where $?a_i \neq t_i$ for all $i$. It interprets every term with respect to the environment, replacing $?a_i$ by $t_i$ on the fly. Whenever the algorithm encounters one of the $?a_i$ it follows the assignment, recursively operating on $t_i$. This terminates: the occurs check prevents cycles like $[(?b, f(?a)), (?a, g(?b))]$.

Most implementations represent variables by pointers for quick updating. To be different, we present a non-destructive version. The environment is represented by a list of assignments: (variable, term) pairs.

Calling `lookup(a,env)` returns the empty list if `a` is not assigned in `env`, and otherwise returns this assignment in a list.

```
fun lookup (X, []) = []
  | lookup (X, (Y,z)::env) = if X = (Y:string) then [z]
                             else  lookup(X,env);
```

Updating means just adding pairs to the list.

## 3.4 The ML code for unification

The unification code has a lot of nested functions. These refer to global variables to minimize parameter passing.

The main function is `unify_terms`. Calling `unify_terms(ts,us,env)` tackles a set of equations given by two lists of terms `ts` and `us`. It fails if their lengths differ. Given two non-empty lists, it calls `unify_term` to unify their heads, then calls itself using the updated environment.

Calling `chasevar` repeatedly replaces its argument, if a variable, by its assignment. For example, if `env` contains $?a \mapsto ?b$ and $?b \mapsto ?c$ then `chasevar` will map $?a$ to $?c$. Doing this 'variable chasing' elsewhere may let the occurs check falsely report that $?c$ occurs in $?a$. Early versions of Folderol contained this error!

Calling `unify_var` unifies a variable with a term. Local functions `occs` and `occsl` (for lists of terms) search for occurrences of the variable. If found, exception `UNIFY` signals the failure, aborting the entire unification. Otherwise `unify_var` returns an updated environment.

Calling `unify_term(t,u)` attempts to unify two terms within the global envi-ronment `env`.

```
exception UNIFY;

fun unify_terms ([],[], env) = env
  | unify_terms (t::ts, u::us, env) =
    let fun chasevar (Var a) =   (*Chase variable assignments*)
              (case  lookup(a,env)  of
                  u::_ => chasevar u  |  [] => Var a)
          | chasevar t = t;
        fun unify_var (Var a, t) = (*unification with var*)
              let fun occs (Fun(_,ts)) = occsl ts
                    | occs (Param(_,bs)) = occsl(map Var bs)
                    | occs (Var b) =  a=b
                                  orelse occsl(lookup(b,env))
                    | occs _ = false
                  and occsl [] = false
                    | occsl(t::ts) = occs t  orelse  occsl ts
              in  if t = Var a  then  env
                  else if occs t then  raise UNIFY else (a,t)::env
              end
          | unify_var (t,u) = unify_term(t,u)
        and unify_term (Var a, t) =
                unify_var (chasevar (Var a), chasevar t)
          | unify_term (t, Var a) =
                unify_var (chasevar (Var a), chasevar t)
          | unify_term (Param(a,_), Param(b,_)) =
                if a=b then env  else  raise UNIFY
          | unify_term (Fun(a,ts), Fun(b,us)) =
                if a=b then unify_terms(ts,us,env) else raise UNIFY
          | unify_term _ =  raise UNIFY
    in  unify_terms (ts, us, unify_term (t,u))  end
  | unify_terms _ =  raise UNIFY;
```

Finally, `unify` handles atomic formulae: $P(t_1, \ldots, t_m)$ and $Q(u_1, \ldots, u_n)$ are unifiable precisely when $P = Q$, $m = n$, and all the pairs of arguments are unifiable.

```
fun unify (Pred(a,ts), Pred(b,us), env) =
        if a=b then unify_terms(ts,us,env)  else   raise UNIFY
  | unify _ =  raise UNIFY;
```

## 3.5 Extensions and omissions

Folderol unifies only atomic formulae. Could we make unification handle quantifiers? We should not 'unify' $\exists x.P(x)$ with $\exists x.P(?a)$ by setting $?a$ to $x$. It is easy to make the occurs check reject terms containing 'loose' bound variables. It is much harder to make unification take account of $\beta$-conversion and find instantiations of function variables. This *higher-order unification* has many interesting applications, but the general problem is undecidable [Huet 1975].

The occurs check is essential in theorem-proving for correct quantifier reasoning. Most Prolog implementations omit the check to gain speed; the price can be circular data structures and looping.

Folderol's unification algorithm is naïve. Though it usually performs well, it requires exponential time in rare cases. There is an efficient algorithm [Martelli and Montanari 1982] that safely omits the occurs check by a sophisticated sorting of the equations. The Martelli/Montanari algorithm is mainly of theoretical importance, but their analysis of unification has had a profound effect on the literature.

## 3.6  Instantiation by the environment

The environment speeds unification by delaying the substitutions. Substitutions can be delayed forever: the environment can be passed from one unification to the next, accumulating ever more assignments. The technique of *structure sharing* even handles variable renaming [Boyer and Moore 1972]. This technique is used in resolution theorem provers and some Prolog systems.

Folderol uses environments only in unification. After a successful unification, it copies out (or *instantiates*) the entire proof, performing all substitutions indicated in the environment. There are two reasons for this. Environments complicate coding, and lookups can be slow.

To instantiate parameters we need to gather all the variables in a term. Folderol uses some general-purpose gathering functionals.

```
fun accumulate f ([], y) = y
  | accumulate f (x::xs, y) = accumulate f (xs, f(x,y));

fun accum_form f (Pred(_,ts),bs) = accumulate f (ts, bs)
  | accum_form f (Conn(_,As),bs) = accumulate(accum_form f)(As,bs)
  | accum_form f (Quant(_,_,A),bs) = accum_form f (A,bs);
```

The function `accumulate` turns a 'gathering function' $f : (\alpha \times \beta) \to \beta$ into one of type $(\alpha\,list \times \beta) \to \beta$: one that, for $n \geq 0$, maps

$$([x_1, x_2, \ldots, x_n], y) \;\longmapsto\; f(x_n, \ldots, f(x_2, f(x_1, y))\ldots)$$

Similarly, `accum_form` turns a gathering function $f : (term \times \beta) \to \beta$ into one of type $(form \times \beta) \to \beta$: one that maps

$$(A, y) \;\longmapsto\; f(t_n, \ldots, f(t_2, f(t_1, y))\ldots)$$

where the terms $[t_1, t_2, \ldots, t_n]$ are all the arguments of predicates in the formula $A$. Gathering functionals uniformly extend a term operation to handle term lists and formulae. They promote brevity and clarity, and are typical of functional programming.

The function `vars_in_term` accumulates the distinct variable names in a term. For a variable it calls `ins` to add the name to the list. For a function application it calls `accumulate` to process the argument list.

```
infix ins;   (*insertion into list if not already there*)
fun x ins xs = if x mem xs then  xs   else   x::xs;

fun vars_in_term (Var a, bs) = a ins bs
  | vars_in_term (Fun(_,ts), bs) = accumulate vars_in_term (ts,bs)
  | vars_in_term (_, bs) = bs;
```

Calling `inst_term env t` instantiates term `t` using environment `env`. Observe how it instantiates the variable $?a$. If the environment has an assignment $?a \mapsto u$ then a recursive call on $u$ is necessary; otherwise the result is $?a$. When `inst_term` receives a parameter $a[?b_1, \ldots, ?b_n]$, the `map(inst_term...)bs` converts the variables into terms. Their variables are gathered into a new list.

```
fun inst_term env (Fun(a,ts)) = Fun(a, map (inst_term env) ts)
  | inst_term env (Param(a,bs)) =
        Param(a, accumulate vars_in_term
                     (map (inst_term env o Var) bs, [])  )
  | inst_term env (Var a) =
      (case  lookup(a,env)  of
          u::_ =>  inst_term env u
        | []   =>  Var a)
  | inst_term env t = t;
```

# 4   Inference in Folderol

Folderol builds a proof upwards from the desired goal. Each inference rule reduces some goal to zero or more subgoals. The proof is complete when no subgoals remain. Each proof step involves several tasks:

- selection of a goal

- selection of an inference rule

- construction of the subgoals

- observing if the new goals are immediately solvable

We have already discussed these tasks. Recall that the goal $\Gamma \vdash \Delta$ is a list of entries, each containing a cost, a side (`Left` or `Right`), and a formula. A proof state, or `goaltable`, is simply a list of goals, though other information could be stored there.

```
datatype side = Left | Right;
type entry = int * side * form;
type goal = entry list;
type goaltable = goal list;
```

## 4.1  Solving a goal

The goal $A_1, \ldots, A_m \models B_1, \ldots, B_n$ is solved by unifying some $A_i$ and $B_j$ — there are $m \times n$ possible combinations. To save time, Folderol considers only atomic formulae; two complicated formulae are unlikely to unify.[3]

The goal is represented as a list of triples, so `split_goal` has the job of producing the two lists $A_1, \ldots, A_m$ and $B_1, \ldots, B_n$. Reversing the input list makes the output lists come out in the correct order. Thus Folderol prefers new formulae to old and is less prone to looping. If Folderol's analysis of formulae were less superficial the order would not matter.

```
fun split_goal G =
  let fun split (As,Bs, []: goal) = (As,Bs)
        | split (As,Bs, (_,Left,A)::H) = split (A::As,Bs, H)
        | split (As,Bs, (_,Right,B)::H) = split (As, B::Bs, H)
  in  split([], [], rev G)  end;
```

In `solve_goal` two nested loops compare every atomic $A_i$ and $B_j$. If unifiable then `findB` returns a 1-element list containing $A_i$ paired with the unifier; these together determine the goal's 'success formula'. The usual outcome is exception `UNIFY` and then trying $B_{j+1}$. After trying all pairs without success, `findA` returns the empty list. If there are several ways of solving a goal then `solve_goal` returns

---

[3]Classifying the formulae, say by predicate name, could speed the process further. Testing a goal with no repeated predicates on either side would require $\min\{m,n\}$ unifications.

the first.

```
fun filter p [] = []
  | filter p (x::xs) = if p(x) then  x :: filter p xs
                       else  filter p xs;

fun is_pred (Pred _) = true  |  is_pred _ = false;

fun solve_goal G =
    let fun findA ([], _) = []     (*failure*)
          | findA (A::As, Bs) =
             let fun findB [] = findA (As,Bs)
                   | findB (B::Bs) = [ (A, unify(A,B,[])) ]
                        handle UNIFY => findB Bs
             in  findB Bs  end
        val (As,Bs) = split_goal G
    in  findA(filter is_pred As, filter is_pred Bs)  end;
```

The function `insert_goals` takes a list of new goals, a list of success formulae (initially empty), and a goaltable. It tries to solve each new goal rather than simply adding it to the goaltable. After solving a goal, `insert_goals` instantiates all the other goals with the resulting environment, since its variables may appear in other goals. If `solve_goal` succeeds, its success formula is accumulated for printing the proof trace.

```
fun insert_goals ([], As, tab) = (As,tab)
  | insert_goals (G::Gs, As, tab) =
      case  solve_goal G  of
         (A,env)::_ =>
            insert_goals (inst_goals env Gs,
                          (inst_form env A) :: As,
                          inst_goals env tab)
       | [] =>  insert_goals (Gs, As, G::tab);
```

## 4.2   Selecting a rule

Folderol selects a rule by looking at the outermost connective of each formula in the selected goal. A rule that produces the fewest subgoals is chosen, but ∀:left and ∃:right are used only if no other rules apply.

Function `cost` does a case analysis on the formula's connective and side (`Left` or `Right`). Function `paircost` attaches a cost to a (side, formula) pair. The cost of

∀:left and ∃:right is three and the cost of other rules is the number of premises.

```
fun cost (_,     Conn("~", _))         = 1   (*a single subgoal*)
  | cost (Left,  Conn("&", _))         = 1
  | cost (Right, Conn("|", _))         = 1
  | cost (Right, Conn("-->", _))       = 1
  | cost (Right, Quant("ALL",_,_))     = 1
  | cost (Left,  Quant("EXISTS",_,_))  = 1
  | cost (Right, Conn("&", _))         = 2   (*2 subgoals*)
  | cost (Left,  Conn("|", _))         = 2
  | cost (Left,  Conn("-->", _))       = 2
  | cost (_    , Conn("<->", _))       = 2
  | cost (Left,  Quant("ALL",_,_))     = 3   (*quant expansion*)
  | cost (Right, Quant("EXISTS",_,_))  = 3   (*quant expansion*)
  | cost _ = 4 ;                             (*no reductions*)

fun paircost (si,A) = (cost(si,A), si, A);
```

The entries in a goal are ordered by cost — the first entry is cheapest. The function `insert` maintains an ordered list given a comparison function `less` for sort keys. If the comparison is < then `insert` places the new entry first among entries of equal cost; if ≤ then the entry goes last. Thus `insert_early` puts the entry where it may be chosen earlier, while `insert_late`'s effect is the opposite.

```
fun insert less =
  let fun insr (x,[]) = [x]
        | insr (x,y::ys) = if less(y,x) then y::insr(x,ys)
                                        else x::y::ys
  in  insr  end;

fun entry_less ((m,_,_): entry, (n,_,_): entry) = m<n;
val insert_early = insert entry_less;

fun entry_lesseq ((m,_,_): entry, (n,_,_): entry) = m<=n;
val insert_late  = insert entry_lesseq;
```

The quantified formula in ∀:left or ∃:right is put last among its fellows so that they will get a turn.

## 4.3  Constructing the subgoals

The sequent calculus LK is suited for backwards proof because it has the *subformula property* — every formula in a proof is a subformula of the original goal. True, quantifier rules ∀:left and ∃:right introduce unknown terms (requiring an unusual notion of subformula), but these are just meta-variables. The key point is this: once we have chosen a rule and a formula from the goal, the subgoals are completely determined.

A goal is a list of triples. A rule is applied to the head of this list. The tail holds the remaining formulae, which must be included in each subgoal. Each subgoal is made from the tail by adding new (side, formula) pairs. Calling `new_goal G pairs` copies the `pairs` into goal `G`, which is the tail of the goal. It calls `paircost` to attach a cost to each new formula.

Since a rule may make more than one subgoal, `new_goals` forms a list of subgoals from a goal and a list of new (side, formula) pairs.

```
fun new_goal G pairs =
    accumulate insert_early (map paircost pairs, G);

fun new_goals G pairslist = map (new_goal G) pairslist;
```

The function `reduce_goal` handles all the rules. Given a formula and its side (`Left` or `Right`), it uses the immediate subformulae to build subgoals. Function `goals` permits a concise description of each subgoal.

For rules ∀:right and ∃:left it generates a fresh parameter and attaches all the variables in the goal. For ∀:left and ∃:right it generates a fresh variable; the subgoal contains the original entry inserted 'late' and the new entry inserted 'early'. Exception `REDUCE` indicates that no reductions are possible, indicating that all the formulae are atomic.

```
exception REDUCE;

fun reduce_goal (pair, G) =
 let val goals = new_goals G;
  fun vars_in A = vars_in_goal (G, vars_in_form(A,[]));
  fun subparam A = subst_bound (Param(gensym(), vars_in A)) A;
  fun subvar A   = subst_bound (Var(gensym())) A;
  fun red(_,Right,Conn("~",[A]))   = goals[[(Left,A)]]
    | red(_,Left, Conn("~",[A]))   = goals[[(Right,A)]]
    | red(_,Right,Conn("&",[A,B])) = goals[[(Right,A)],[(Right,B)]]
    | red(_,Left, Conn("&",[A,B])) = goals[[(Left,A),(Left,B)]]
    | red(_,Right,Conn("|",[A,B])) = goals[[(Right,A),(Right,B)]]
    | red(_,Left, Conn("|",[A,B])) = goals[[(Left,A)],[(Left,B)]]
    | red(_,Right,Conn("-->",[A,B]))=goals[[(Left,A),(Right,B)]]
    | red(_,Left, Conn("-->",[A,B]))=goals[[(Right,A)],[(Left,B)]]
    | red(_,Right,Conn("<->",[A,B])) =
          goals[[(Left,A),(Right,B)],[(Right,A),(Left,B)]]
    | red(_,Left, Conn("<->",[A,B])) =
          goals[[(Left,A),(Left,B)],[(Right,A),(Right,B)]]
    | red(_,Right,Quant("ALL",_,A)) = goals[[(Right, subparam A)]]
    | red(_,Left, Quant("ALL",_,A)) =
        [ insert_early (paircost(Left, subvar A),
                        insert_late(pair,G)) ]
    | red(_,Right,Quant("EXISTS",_,A)) =
        [ insert_early (paircost(Right, subvar A),
                        insert_late(pair,G)) ]
    | red(_,Left,Quant("EXISTS",_,A)) = goals[[(Left, subparam A)]]
    | red _ = raise REDUCE
 in  red pair  end;
```

The function `gensym`, like its LISP namesake, generates unique variable names. Functional programmers may hear with dismay that `gensym` increments a counter. We could do without the reference variable `varcount` by keeping the count with the goal list, but why? Side-effects are more dangerous in programmable theorem provers like LCF and Isabelle, where a proof strategy may spawn subproofs that must run independently. Yet LCF's simplifier uses a form of `gensym`.

Up to this point Folderol is largely applicative. As we approach the top-level commands it becomes more and more imperative. Applying a rule prints tracing information as a side effect, and the proof state is kept in reference variables.

## 4.4  Goal selection

A proof fails if some goal is unprovable, so a theorem prover should choose the goal that seems most likely to fail. Folderol is not intended to detect unprovability, so it always chooses the next goal using 'last in first out': like a stack. Folderol keeps a list of goals (the goaltable). At each step it replaces the first goal by its subgoals. This simple structure makes proofs easier to follow.

The function `proof_step` gives the head of the goal list to `reduce_goal`, and then `insert_goals` creates a new goaltable. Calling `proof_steps` with $n \geq 0$ performs up to $n$ steps; negative $n$ allows unbounded repetition. These functions print a proof trace.

```
fun proof_step [] = [] : goaltable
  | proof_step ([]::tab) = raise ERROR "Empty goal"
  | proof_step ((ent::G)::tab) =
      let val (As,newtab) = insert_goals(reduce_goal(ent,G),[],tab)
      in  print_step(ent,tab,As);   newtab   end;

fun proof_steps (_,[]) = []      (*success -- no goals*)
  | proof_steps (0,tab) = tab
  | proof_steps (n,tab) = proof_steps (n-1, proof_step tab)
      handle REDUCE => (prints"\n**No proof rules applicable**\n";
                       tab);
```

Let us skip the code for reading and printing goaltables. At the end of the program are Folderol's commands:

- to read a goal, a sequent $\Gamma \models \Delta$ or formula $\models B$
- to perform one step, or $n$ steps, or run without bound
- to read a goal and immediately run

Folderol has a top-level proof state, and this requires a reference variable, `the_goaltable`. The only other reference is the variable counter for `gensym`.

```
val the_goaltable = ref ([] : goaltable);

fun set_tab tab = (the_goaltable := tab;  print_tab tab);

fun read_goalseq (Astrs,Bstrs) =
    (init_gensym();   set_tab(read_tab (Astrs,Bstrs)));

fun read_goal Bstr = read_goalseq([],[Bstr]);

fun step()  = set_tab (proof_step(!the_goaltable));
fun steps n = set_tab (proof_steps (max(n,0), !the_goaltable));
fun run()   = set_tab (proof_steps (~1, !the_goaltable));

fun run_goalseq (Astrs,Bstrs) = (read_goalseq(Astrs,Bstrs); run());
fun run_goal b = run_goalseq([],[b]);
```

Note that Folderol may reorder the formulae in your original goal, so you may not recognize it.

# 5 Folderol in action

The computer sessions below were run using D. C. J. Matthews's Poly/ML, a compiler for Standard ML. Input lines to Folderol begin with the Poly/ML prompt characters (`>` or `#`). All other lines are output. I have edited the output to make it more compact and readable.

## 5.1 Propositional examples

Pelletier [1986] has published a list of graded problems in classical first-order logic for testing theorem provers. Folderol, running on a Sun-3,[4] can prove any of the propositional ones in 0.1 seconds. Folderol is complete for propositional logic.

### 5.1.1 A distributive law

Here is part of a distributive law:

$$(P \vee Q) \wedge (P \vee R) \rightarrow P \vee (Q \wedge R)$$

The command `read_goal` accepts this formula; then `step()` performs the only possible rule. Observe how Folderol reports its choice of rule, here →:right.

```
> read_goal "(P | Q) & (P | R)  -->  P | (Q & R)";
empty   |-  (P | Q) & (P | R) --> P | Q & R

> step();
-->:right
(P | Q) & (P | R)   |-  P | (Q & R)
```

Reducing the conjunction on the left or the disjunction on the right will produce one subgoal. A goal is a single ordered list but Folderol prints it as two lists, the left and right formulae. We can only be sure that Folderol will reduce the first formula on the left or the right. It happens that Folderol first reduces the disjunction. Next it ignores the new conjunction on the right (which would produce two subgoals) and reduces the conjunction on the left.

```
> step();
|:right
(P | Q) & (P | R)   |-  Q & R, P

> step();
&:left
P | R, P | Q  |-  Q & R, P
```

Now all three possible reductions will make two subgoals. Arbitrarily the $P \vee R$ is reduced first. The $P$ subgoal is solved (hence the P in the trace) while the $R$ subgoal remains. Next $P \lor Q$ behaves similarly. Finally $Q \land R$, on the right, is split when $Q$ and $R$ are both assumed true. Thus both subgoals are solved and the proof is complete.

---

[4]Sun-3 is a trade mark of Sun Microsystems, Inc.

```
> step();
|:left    P
P | Q, R  |-  Q & R, P

> step();
|:left    P
Q, R  |-  Q & R, P

> step();
&:right   Q    R
No more goals: proof finished
```

### 5.1.2   An associative law for ↔

The classical identity
$$((P \leftrightarrow Q) \leftrightarrow R) \leftrightarrow (P \leftrightarrow (Q \leftrightarrow R)),$$

one of Pelletier's sample problems, is excellent for illustrating the ↔ rules.

To shorten the proof let us consider only one direction. The command `read_goalseq` accepts this as a sequent. Since both ↔ rules have two premises, Folderol's initial choice of ↔:right is arbitrary. The next step removes $Q \leftrightarrow R$ on the left.

```
> read_goalseq (["((P <-> Q) <-> R)"],  ["(P <-> (Q <-> R))"]);
(P <-> Q) <-> R  |-  P <-> (Q <-> R)

> step();
<->:right
Q <-> R, (P <-> Q) <-> R  |-  P
(P <-> Q) <-> R, P  |-  Q <-> R

> step();
 <->:left
(P <-> Q) <-> R  |-  R, Q, P
(P <-> Q) <-> R, R, Q  |-  P
(P <-> Q) <-> R, P  |-  Q <-> R
```

Reducing $(P \leftrightarrow Q) \leftrightarrow R$ has the effect of replacing $R$ by $P \leftrightarrow Q$; the nontrivial subgoal has $P \leftrightarrow Q$, $Q$, and $P$ on the right.[5] Note that if $Q$ and $P$ are both false then $P \leftrightarrow Q$ is true. So applying $\leftrightarrow$:right completely solves this subgoal.

---

[5] It also has two copies of $R$; Folderol ought to remove duplicate formulae and other redundancies.

```
> step();
  <->:left    R
empty  |-  P <-> Q, R, R, Q, P
(P <-> Q) <-> R, R, Q  |-  P
(P <-> Q) <-> R, P  |-  Q <-> R

> step();
  <->:right    P    Q
(P <-> Q) <-> R, R, Q  |-  P
(P <-> Q) <-> R, P  |-  Q <-> R
```

Here we informally see that assuming $(P \leftrightarrow Q) \leftrightarrow R$ and $R$ amounts to assuming $P \leftrightarrow Q$, as indeed happens. In the next step, assuming $P \leftrightarrow Q$ and $Q$ amounts to assuming $P$. Impatience now suggests typing `run()` to finish.

```
> step();
 <->:left    R
P <-> Q, R, R, Q  |-  P
(P <-> Q) <-> R, P  |-  Q <-> R

> step();
 <->:left    P    Q
(P <-> Q) <-> R, P  |-  Q <-> R

> run();
<->:right
 <->:left    R
 <->:left    Q    P
<->:left    R
<->:right    Q    P
No more goals: proof finished
```

Observe how the indentation of the proof trace varies. Folderol prints additional space before the name of a rule to indicate the number of subgoals. Often the indentation will increase for a time and then decrease; if it gets bigger and bigger, look out!

### 5.1.3 The completeness of propositional logic

Folderol is complete for propositional logic. For a formula $A$, if $A$ is valid then it constructs a proof; if $A$ is invalid then it constructs a model that falsifies $A$ [Gallier 1986, page 71].

Is this formula valid?

$$(P \rightarrow (Q \rightarrow R)) \rightarrow (P \vee Q \rightarrow R)$$

Folderol answers in no time.

```
> run_goal "(P --> (Q-->R))  -->  (P | Q --> R)";
empty  |-  (P --> (Q --> R)) --> (P | Q --> R)

-->:right
-->:right
|:left
 -->:left
  -->:left   Q   R

**No proof rules applicable**
Q   |-   P, R
P --> (Q --> R), P  |-  R
```

The first goal is not provable: if $Q$ is true while $P$ and $R$ are false then the goal is false, as is the original goal.

This session demonstrates the power of decidability: every question can be answered.

## 5.2  Quantifier examples

Most of the data structures were designed around quantifiers. Let us verify that parameters and variables really work for the examples we discussed originally. Then we shall see a complicated theorem Folderol can prove — and a simple one it cannot prove.

### 5.2.1  Valid reasoning permitted

Here is one of our quantifier examples:

$$\forall x.R(x,x) \vdash \forall x.\exists y.R(x,y)$$

We enter the goal. The first step is $\forall$:right since the alternative, $\forall$:left, has high cost. Folderol prints a table of parameters showing the associated variables; the parameter a has none.

```
> read_goalseq ( ["ALL x.R(x,x)"],  ["ALL x. EXISTS y. R(x,y)"] ) ;
ALL x. R(x,x)  |-  ALL x. EXISTS y. R(x,y)

> step();
ALL:right
ALL x. R(x,x)  |-  EXISTS y. R(a,y)

Param     Not allowed in
a
```

Now ∃:right is arbitrarily chosen, introducing the variable `?b`. The quantified formula $\exists y.R(a,y)$ is put far back in the queue. So the other quantifier is expanded next; ∀:left adds the assumption $R(?c,?c)$, which immediately unifies with $R(a,?b)$ giving $R(a,a)$.

```
> step();
EXISTS:right
ALL x. R(x,x)  |-  EXISTS y. R(a,y), R(a,?b)

> step();
ALL:left   R(a,a)
No more goals: proof finished
```

### 5.2.2 Invalid reasoning forbidden

Reversing the quantifiers turns the last example into an invalid sequent:

$$\forall x.R(x,x) \models \exists y.\forall x.R(x,y)$$

Given this goal Folderol arbitrarily chooses ∃:right, introducing `?a`.

```
> read_goalseq (["ALL x.R(x,x)"],  ["EXISTS y. ALL x. R(x,y)"]) ;
ALL x. R(x,x)  |-  EXISTS y. ALL x. R(x,y)

> step();
EXISTS:right
ALL x. R(x,x)  |-  ALL x. R(x,?a), EXISTS y. ALL x. R(x,y)
```

The new formula has the least cost; ∀:right introduces the parameter `b`, depending on `?a`. Next it is time to expand $\forall x.R(x,x)$.

```
> step();
ALL:right
ALL x. R(x,x)  |-  EXISTS y. ALL x. R(x,y), R(b,?a)

Param     Not allowed in
b         (?a)

> step();
ALL:left
ALL x. R(x,x), R(?c,?c)  |-  EXISTS y. ALL x. R(x,y), R(b,?a)

Param     Not allowed in
b         (?a)
```

Folderol has left this subgoal because $R(?c,?c)$ and $R(b,?a)$ are not unifiable. They may look unifiable, but assigning $?a \mapsto b[?a]$ would be circular. Instead we expand the quantifiers again. The next application of $\forall$:right adds $R(e, ?d)$, and $?d \mapsto e[?a, ?c, ?d]$ is also circular. Thus $R(e, ?d)$ is not unifiable with $R(?c, ?c)$, nor with the new assumption $R(?f, ?f)$.

```
> step();
EXISTS:right
ALL x. R(x,x), R(?c,?c)
|-  ALL x. R(x,?d), EXISTS y. ALL x. R(x,y), R(b,?a)

Param     Not allowed in
b         (?a)

> step();
ALL:right
ALL x. R(x,x), R(?c,?c)
|-  EXISTS y. ALL x. R(x,y), R(e,?d), R(b,?a)

Param     Not allowed in
b         (?a)
e         (?a,?c,?d)

> step();
ALL:left
ALL x. R(x,x), R(?f,?f), R(?c,?c)
|-  EXISTS y. ALL x. R(x,y), R(e,?d), R(b,?a)

Param     Not allowed in
b         (?a)
e         (?a,?c,?d)
```

Obviously we are getting nowhere. But this is less obvious to Folderol.

```
> steps 9;
EXISTS:right
ALL:right
ALL:left
EXISTS:right
ALL:right
ALL:left
EXISTS:right
ALL:right
ALL:left

ALL x. R(x,x), R(?o,?o), R(?l,?l), R(?i,?i), R(?f,?f), R(?c,?c)
|-  EXISTS y. ALL x. R(x,y), R(n,?m), R(k,?j),
    R(h,?g), R(e,?d), R(b,?a)

Param     Not allowed in
b         (?a)
e         (?a,?c,?d)
h         (?a,?c,?d,?f,?g)
k         (?a,?c,?d,?f,?g,?i,?j)
n         (?a,?c,?d,?f,?g,?i,?j,?l,?m)
```

Folderol will never quit. The general problem of when to quit is undecidable.

### 5.2.3 A complicated proof

Pelletier's problem 29 dates back to *Principia Mathematica* (*11.71):

$$(\exists x.P(x)) \wedge (\exists x.Q(x)) \;\vdash (\forall x.P(x) \to R(x)) \wedge (\forall x.Q(x) \to S(x))$$
$$\leftrightarrow (\forall xy.P(x) \wedge Q(y) \to R(x) \wedge S(y))$$

Let us enter this sequent to Folderol:

```
> read_goalseq (["(EXISTS x. P(x)) & (EXISTS x. Q(x))"],
#               ["(ALL x. P(x)-->R(x)) & (ALL x. Q(x)-->S(x)) <->  \
#\                (ALL x. ALL y. P(x) & Q(y) --> R(x) & S(y))"]);

(EXISTS x. P(x)) & (EXISTS x. Q(x))
|-  (ALL x. P(x) --> R(x)) & (ALL x. Q(x) --> S(x))
    <-> (ALL x. ALL y. P(x) & Q(y) --> R(x) & S(y))
```

The proof begins with routine reductions on the left followed by analysis of one direction of the equivalence. The parameters a, b, c depend on no variables;

redundant parameter listings are omitted.

```
> steps 7;
&:left
EXISTS:left
EXISTS:left
<->:right
 &:right
   ALL:right
   -->:right

ALL x. ALL y. P(x) & Q(y) --> R(x) & S(y), Q(c), P(b), Q(a)
|-  S(c)

ALL x. ALL y. P(x) & Q(y) --> R(x) & S(y), P(b), Q(a)
|-  ALL x. P(x) --> R(x)

(ALL x. P(x) --> R(x)) & (ALL x. Q(x) --> S(x)), P(b), Q(a)
|-  ALL x. ALL y. P(x) & Q(y) --> R(x) & S(y)
```

The first subgoal involves proving $S(c)$ from $Q(c)$ and $P(b)$ using the quantified formula. This takes five steps.

```
> steps 5;
  ALL:left
  ALL:left
  -->:left
   &:left   S(c)
   &:right  P(b)   Q(c)

ALL x. ALL y. P(x) & Q(y) --> R(x) & S(y), P(b), Q(a)
|-  ALL x. P(x) --> R(x)

(ALL x. P(x) --> R(x)) & (ALL x. Q(x) --> S(x)), P(b), Q(a)
|-  ALL x. ALL y. P(x) & Q(y) --> R(x) & S(y)
```

Two reductions on the right produce an analogous situation.

```
> steps 2;
 ALL:right
 -->:right

ALL x. ALL y. P(x) & Q(y) --> R(x) & S(y), P(f), P(b), Q(a)
|-  R(f)

(ALL x. P(x) --> R(x)) & (ALL x. Q(x) --> S(x)), P(b), Q(a)
|-  ALL x. ALL y. P(x) & Q(y) --> R(x) & S(y)
```

We now prove $R(f)$ from $P(f)$ and $Q(a)$:

```
> steps 5;
 ALL:left
 ALL:left
 -->:left
  &:left    R(f)
 &:right    P(f)    Q(a)

(ALL x. P(x) --> R(x)) & (ALL x. Q(x) --> S(x)), P(b), Q(a)
|-  ALL x. ALL y. P(x) & Q(y) --> R(x) & S(y)
```

One direction of the original goal is proved. Now we tackle the other direction.

```
> steps 6;
ALL:right
ALL:right
-->:right
&:left
&:left
&:right

ALL x. Q(x) --> S(x), ALL x. P(x) --> R(x), Q(j), P(i), P(b), Q(a)
|-  S(j)

ALL x. Q(x) --> S(x), ALL x. P(x) --> R(x), Q(j), P(i), P(b), Q(a)
|-  R(i)
```

The two goals are similar. The first Folderol proves directly.

```
> steps 2;
 ALL:left
 -->:left    Q(j)    S(j)

ALL x. Q(x) --> S(x), ALL x. P(x) --> R(x), Q(j), P(i), P(b), Q(a)
|-  R(i)
```

But here Folderol fails to see that $\forall x. Q(x) \to S(x)$ is irrelevant. Luckily the proof succeeds when the correct formula is picked.

```
> steps 2;
ALL:left
-->:left    Q(j)

ALL x. P(x) --> R(x), ALL x. Q(x) --> S(x),
            S(j), Q(j), P(i), P(b), Q(a)    |-  R(i)

> steps 2;
ALL:left
-->:left    P(i)   R(i)
No more goals: proof finished
```

### 5.2.4   Folderol fooled

In the last example, reducing the wrong quantifier did no harm. But needless quantifier expansions can cause subgoals to multiply: reducing $\forall z.A \vee B$ on the left causes a case split. Folderol expands quantifiers and instantiates variables according to routines that never consider the proof as a whole.

Folderol easily proves the following contrapositive:

$$\forall x.P(x) \to Q(x) \models \forall x.\neg Q(x) \to \neg P(x)$$

Adding $\exists x.P(x)$ to this yields a more complicated theorem:

$$\exists x.P(x), \forall x.P(x) \to Q(x) \models \forall x.\neg Q(x) \to \neg P(x)$$

The extra formula distracts Folderol from the correct instantiation.

```
> read_goalseq (["EXISTS x. P(x)", "ALL x. P(x)-->Q(x)"],
#               ["ALL x. ~Q(x) --> ~P(x)"]);

EXISTS x. P(x), ALL x. P(x) --> Q(x)   |-   ALL x. ~Q(x) --> ~P(x)

> steps 5;
ALL:right
-->:right
~:right
~:left
EXISTS:left

ALL x. P(x) --> Q(x), P(b), P(a)   |-   Q(a)
```

The proof should follow by putting $a$ into $\forall x.P(x) \to Q(x)$ and using $P(a)$. The $P(b)$ is simply noise.

```
> steps 6;
ALL:left
-->:left    P(b)
ALL:left
-->:left    P(b)
ALL:left
-->:left    P(b)

ALL x. P(x) --> Q(x), Q(b), Q(b), Q(b), P(b), P(a)   |-  Q(a)
```

Reducing $\forall x.P(x) \rightarrow Q(x)$ produces the subgoal $P(b), P(a) \models P(?c)$. Folderol always chooses the first solution, $b$, instead of $a$, and gets nowhere.

In another version of Folderol, `solve_goal` chooses among multiple solutions randomly. When lucky (!) it works well, but its performance is not reliable.

## 5.3  Beyond Folderol: advanced automatic methods

Folderol is a toy. Its purpose is to illustrate coding techniques; its main strength is its simplicity. Naïve proof methods have surprising power but obvious limitations. How can we do better?

Function `solve_goal` always takes the first solution it finds, though this may block the proof of other goals. Sometimes the prover should choose another solution or even ignore them all (thus leaving the goal open to further proof steps). Backtracking is the obvious way to search these possibilities. I have written such a prover in Prolog.

The predicate `proof(As,Bs,N,P)` succeeds if it proves the sequent $As \models Bs$, where `N` limits quantifier expansions and `P` is the proof tree. The limit on quantifier expansion (applications of $\forall$:left and $\exists$:right) makes the search space finite so that backtracking can explore it. The program returns a proof tree instead of tracing its search.

The Prolog prover is 2.5 times smaller than the ML one. It does not need a parser because of Prolog operator declarations. Within a goal, sorting the formulae by cost is no longer necessary. Instead the rules are tried in order of cost. Here are the clauses for the rules ¬:right, ∧:left, and ∧:right. Predicate `delmem` searches for a given element in a list and deletes it. The cuts prevent backtracking over the choice of rule.

```
delmem(X, [X|Xs], Xs).
delmem(X, [Y|Ys], [Y|Zs]) :- delmem(X, Ys, Zs).

proof(As, Bs, N, notr(P)) :- delmem(~B, Bs, Ds),
        !, proof([B|As], Ds, N, P).

proof(As, Bs, N, andl(P)) :- delmem(A1&A2, As, Cs),
        !, proof([A1,A2|Cs], Bs, N, P).

proof(As, Bs, N, andr(P1,P2)) :- delmem(B1&B2, Bs, Ds),
        !, proof(As, [B1|Ds], N, P1), proof(As, [B2|Ds], N, P2).
```

Predicates `abstract` and `subst` handle abstraction and substitution in the de Bruijn representation. The rule ∀:right is

```
proof(As, Bs, N, allr(abs(Id,P))) :-
        delmem(all(abs(Id,B1)), Bs, Ds),
        make_param((As,Bs), X), !, subst(X,B1,BX),
        proof(As, [BX|Ds], N, P1), abstract(X,P1,P).
```

Finally, if no other rules apply and `N` is positive, then all top-level quantifiers are expanded.

Prolog can be extremely useful for experiments with theorem proving. But the time may come when you are constantly fighting Prolog's view of data and control. To check whether some left side formula is unifiable with some right side formula requires a proper unification predicate, with the occurs check and de Bruijn indices. The code for this predicate is a collection of Prolog hacks — not Logic Programming.

The Prolog version is complete. Given a large enough value of `N` it should find a proof if there is one. Yet on hard problems it is little better than Folderol. Even with `N=2`, the 'finite' search space is astronomical.[6] Both provers are weak because they choose a formula simply by its outer connective, not for its contents or relevance.

The links in a *connection graph* help to decide which formula to reduce next. Kowalski [1975] invented connections for resolution theorem provers (see also [Eisinger 1986]). The graph links pairs of atomic formulae that are potentially *complementary*: unifiable and of opposite signs. With resolution's clause form, the sign of an atomic formula is obvious. With sequents, the sign is determined by counting the surrounding negations (including → and ↔). Observe that applying sequent calculus rules to a formula breaks it apart, bringing its subformulae to the surface; a subformula's sign indicates whether it will end up on the left or the right. Unification must allow for quantified variables using methods such as Folderol's.

The theorem prover HARP [Oppacher and Suen 1988] resembles Folderol in its logical formalism. It uses semantic tableax, which are equivalent to sequents.

---

[6]Of Pelletier's problems 1–45, Folderol failed on 34, 38, and 41–44, while the Prolog version failed on 34, 37, 38, 43, and 45.

HARP is much more powerful than Folderol by virtue of its heuristics and connection methods. But a collection of heuristics is hard to analyze scientifically.

The matrix methods of [Bibel 1987] define a notion of *path* in a formula. A formula is a theorem if all paths within it contain a connection. The number of paths is exponential but each connection rules out a whole class of paths. Matrix methods can determine that a proof can be constructed without constructing one, avoiding certain redundancies in the search.

A specialized but established method is classical *resolution* [Chang and Lee 1973]. To prove $A$, translate $\neg A$ into a set of clauses, each a disjunction of atomic formulae or their negations. Each resolution step takes two clauses and yields a new one. The method succeeds if it produces the empty clause (a contradiction), thus refuting $\neg A$. Because resolution is the only inference rule, the method is easy to analyze, and many refinements have been found. One such refinement led to the language Prolog. Conversely, Prolog systems have been used to make fast resolution provers [Stickel 1988]. Resolution's chief drawback is its clause form, which renders a formula unintelligible.

What about non-classical logics? Each must be treated individually. Intuitionistic logic differs from classical logic by the lack of the one rule $\frac{\neg\neg A}{A}$, but it requires a completely different approach. TPS proves theorems in higher-order logic [Andrews et al. 1984]. It uses general matings (essentially Bibel's method) and higher-order unification. Lincoln Wallen [1990] has developed matrix methods for several modal logics and intuitionistic logic through a careful analysis of their semantics.

A remarkable system exploits an efficient decision procedure for a propositional temporal logic [Clarke, Emerson, and Sistla 1986]. Statements are verified by checking a small number of cases. This is really model checking, not theorem proving, but such an effective method must come under the heading of Algorithmic Proof. They chose their temporal logic mainly for its fast decision procedure: expressiveness had to be secondary.

Powerful automatic techniques are often brittle, unable to accept the slightest change in the logic. Adding an induction rule would count as a revolutionary change. So we have to direct the proof. We next shall see how interactive systems are organized and controlled.

# 6 Interactive theorem proving

Suppose we want to prove something from the transitivity axiom

$$\forall xyz \,.\, x = y \land y = z \to x = z$$

Folderol will apply $\forall$:left to get

$$\forall yz \,.\, ?a = y \land y = z \to ?a = z$$

At the next round, Folderol will apply $\forall$:left to *both* quantified formulae, producing

$$\forall yz \,.\, ?a_1 = y \land y = z \to ?a_1 = z$$

$$\forall z \,.\, ?a = ?b \land ?b = z \to ?a = z$$

Folderol keeps all of these; the next round produces

$$\forall yz \,.\, ?a_2 = y \land y = z \to ?a_2 = z$$

$$\forall z \,.\, ?a_1 = ?b_1 \land ?b_1 = z \to ?a_1 = z$$

$$?a = ?b \land ?b = ?c \to ?a = ?c$$

The last formula may be useful; the others are rapidly multiplying junk.

It is not hard to improve the treatment of nested quantifiers. But the improved prover will still flounder because of transitivity itself. Given the goal

$$a = b, b = a, b = c \models a = c$$

unguided use of transitivity could generate $a = a$, $b = b$, and other useless facts before succeeding. Effective reasoning about equality requires an algorithm specific to the problem at hand. *Rewriting* reduces an expression to normal form. *Equational unification* solves a set of equations in the presence of equational laws. *Congruence closure*, given a set of equations, efficiently tests whether certain other equations hold [Gallier 1986].

The ideal interactive theorem prover would provide all useful algorithms known. The Boyer/Moore theorem prover knows a good many algorithms and how to use them. But new algorithms are constantly being developed and some users would like to invent their own. Equality is just one example of the limitations of general algorithms. Knowledge of the problem domain may suggest effective specialized algorithms. *Tactics* and *tacticals* are a flexible language for describing proofs at a high level.

## 6.1   The Boyer/Moore theorem prover

The control language of the [Boyer and Moore 1979] theorem prover is extremely simple. You can define new data structures and functions, or ask it to prove a theorem. It applies a vast battery of heuristics to theorems it already has. You guide this process by carefully planning a sequence of lemmas, leading to the main theorem. The prover should be told what each lemma is for — rewriting, generalization, induction — or, sometimes, to 'forget' a lemma.

Its success is due to an effective combination of quantifier-free logic and well-founded induction. No quantifiers means no worries about bound variables and unification. Well-founded induction permits proofs that would otherwise involve quantifiers. Consider proving $\forall n \,.\, P(m, n)$ by mathematical induction on $m$. The inductive step ($m > 0$) is to prove $\forall n \,.\, P(m, n)$ from $\forall n \,.\, P(m - 1, n)$. If the proof uses this induction hypothesis for several values of $n$, then the $\forall n$ seems essential. But it is equivalent to prove $P(m, n)$ by well-founded induction on $(m, n)$ under the following well-founded relation:

$$(m', n') \prec (m, n) \quad \text{if and only if} \quad m' < m$$

Then the inductive step is to prove $P(m, n)$ assuming $P(m', n')$ for all $(m', n') \prec (m, n)$. Argue by cases. If $m = 0$ the argument is the same as before; if $m > 0$ then $(m - 1, n') \prec (m, n)$ so we may assume $P(m - 1, n')$ for all $n'$. Thus the previous proof goes through without quantifiers.

Boyer and Moore's treatment of induction is particularly impressive. A well-founded relation is used to prove termination for each recursive function definition. The prover exploits this information to choose a form of induction appropriate for a goal involving several recursive functions. It also knows that proving something by induction may require proving something stronger. It can strengthen a goal by generalizing it or discarding useless information.

One drawback is that the user needs to understand the heuristics and their many interactions. The lack of quantifiers impairs the expressiveness of the logic, despite recent work on bounded quantification [Boyer and Moore 1988].

The prover has done many proofs in pure mathematics, including Gödel's Incompleteness Theorem. A complete computer system has been verified, both software and hardware, from a compiler down to gate level [Bevier et al. 1989].

## 6.2 The Automath languages

The Automath project [de Bruijn 1980] has tackled a different problem: that of expressing mathematical concepts formally. First-order logic is a rich language for expressing statements, but it includes nothing to make statements about. Defining even the natural numbers in first-order logic is not easy. The induction axiom is not first-order unless it is given as an axiom scheme (an infinite set of axioms). While the Boyer/Moore logic includes induction, it is extremely constructive and seems to rule out many forms of classical mathematical reasoning: say, defining the real numbers as certain sets of rationals.

Automath introduced a formal language in which mathematical notions could be assumed or constructed, and even the proofs formalized. The language (actually several were developed) was an extension of the typed $\lambda$-calculus. It provided just enough structure to permit assumptions to be made or discharged. Through this could be defined the basic connectives of logic — then perhaps the natural numbers, the integers, the rationals, and the real numbers. The translation into Automath of Landau's textbook *The Foundations of Analysis* was a triumph that has seldom been surpassed [Jutting 1977].

Automath did not stress interaction; a mathematician would write a 'book' in the language, then submit it to be checked. In another sense, however, Automath was revolutionary. It viewed the introduction of an assumption as the introduction of a variable, and viewed the discharge of an assumption as the creation of a function. It accomplished this using $\Pi$ types (sometimes called 'dependent types'), interpreting propositions as types. These concepts may be familiar now, but Automath was hardly appreciated in its day.

The *Calculus of Constructions*, by Coquand and Huet, continues this work [Formel 1989]. The Calculus is closely related to the Automath languages. Coquand has investigated its formal properties. Huet and others have implemented the Calculus and formalized large pieces of mathematics: the Schroeder-Bernstein Theorem,[7] to name one example. The Calculus is also being applied towards the synthesis of correct programs by proof.

## 6.3 LCF, a programmable theorem prover

Edinburgh LCF was developed during the 1970's by Robin Milner and his colleagues [Gordon, Milner, and Wadsworth 1979]. Its techniques have been adopted in numerous theorem provers.

In order to be extensible, Edinburgh LCF was programmable. (Its Meta Language, called ML, was the predecessor of Standard ML.) The user could write ML functions to process terms, formulae, and theorems. Theorems were not simply created, but proved; they belonged to an *abstract type* called `thm`, which provided the inference rules as functions. Type checking ensured that theorems were only proved by applying rules to axioms and other theorems.

This collection of rule functions is best viewed as a virtual machine code. Like machine instructions they are hard to use directly; they should be used to implement higher levels of abstraction. LCF uses various 'blocks' — *conversions* for rewriting[8] and *tactics* for backwards proof — together with 'mortar' for putting the blocks together. The blocks are typically functions and they are combined by functionals (higher-order functions). Tactics are combined by *tacticals*.

Each tactic specifies a backwards proof step, reducing a goal to subgoals. Most tactics only accept a certain set of goals and *fail* on all goals outside this domain. A tactic succeeds on a goal if it returns any number of subgoals, and proves it if this number is zero. Tactics may also be applied to subgoals, their subgoals, etc., and once they have all been proved, some mechanism returns the desired theorem. Tactics include primitive tactics and compound tactics built using tacticals such as `THEN`, `ORELSE`, `REPEAT`.

- Every rule that may be useful in backwards proof has a corresponding primitive tactic.

- The tactic `tac1 THEN tac2` applies `tac1` to a goal then applies `tac2` to the result. The effect is that of applying both tactics in succession.

- The tactic `tac1 ORELSE tac2` applies `tac1` to a goal. If `tac1` fails (because the goal is outside its domain) then it tries `tac2` as an alternative. If both fail then `ORELSE` fails. Its domain is the union of the domains of `tac1` and `tac2`.

- The tactic `REPEAT tac` applies `tac` repeatedly to a goal.

Ideally, tactics should capture the control structures people use when describing proofs. In practice, tactics do not always work at such a high level. But through tacticals like `REPEAT`, a single user command can perform hundreds of inferences.

---

[7] If $|A| \leq |B|$ and $|B| \leq |A|$ then $|A| = |B|$, where $|\cdots|$ denotes the cardinality of a set.

[8] Conversions were introduced in Cambridge LCF, a derivative of Edinburgh LCF.

Many different conceptions of tactic exist, including those of Edinburgh LCF and Isabelle. They are all consistent with the general ideas above.

## 6.4 Validation-based tactics

An LCF goal is a description of the desired theorem, possibly with additional information.[9] A tactic maps a goal to a list of subgoals and a proof function, which will be used to map a theorem list to a theorem.

```
type proof = thm list -> thm;
type goal = form list * form;
type tactic = goal -> ((goal list) * proof);
```

The proof function is also called a *validation.* It is not the name of a rule but an arbitrary function from theorems to a theorem. Likewise the tactic may use an arbitrary function from the goal to the subgoals. Subgoal construction and proof construction are completely separate.

Relating these operations are the concepts of *achievement* and *validity* [Milner 1985]. Each goal defines some set of theorems that achieve it. If the goal is false then this set is empty. A tactic is *valid* provided: whenever it reduces goal $G$ to subgoals $G_1$, ..., $G_n$ and proof $P$, and theorems $T_1$, ..., $T_n$ achieve $G_1$, ..., $G_n$, then theorem $P[T_1, \ldots, T_n]$ achieves $G$.

In this model, tactical proof has two phases. In the top-down phase, the original goal is decomposed into subgoals, and all are ultimately solved. Then the bottom-up phase, which should be automatic, applies validations to theorems. When a subgoal is solved outright (the tactic returns an empty goal list) the validation is applied to the empty theorem list and returns a theorem. These theorems are given to the validations one level up, and so forth until we reach the root.

Below is a goal tree and the corresponding tree of theorems. We must keep track of the validations and use them to make the $T_i$. Cambridge LCF has a 'subgoal package' that stores validations on a stack and applies them automatically. This package constrains somewhat the selection of the next subgoal.

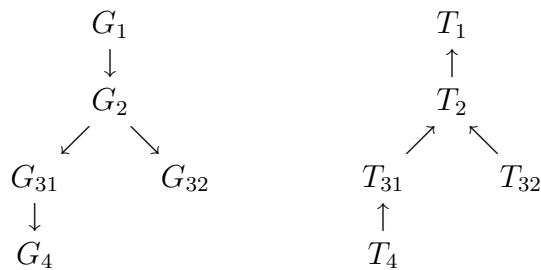

---

[9]In Edinburgh LCF each goal contained a *simpset*, consisting of simplification data; these were dropped in Cambridge LCF.

If all the tactics are primitive, this might be the goal tree for the following proof:

$$\cfrac{\cfrac{\cfrac{A,B \vdash B}{A \vdash B \to B} \;\; \to\text{:right} \quad\quad A \vdash A}{A \vdash (B \to B) \land A} \;\; \land\text{:right}}{\vdash A \to (B \to B) \land A} \;\; \to\text{:right}$$

There should be primitive tactics for all the inference rules. If the rule maps premises $T_1, \ldots, T_n$ to conclusion $T$, the tactic recognises a goal corresponding to $T$ and returns subgoals corresponding to $T_1, \ldots, T_n$, along with a validation that invokes the rule. Sometimes a rule function depends on extra arguments, which the validation must supply. The tactic can usually obtain this information from the goal. If construction of the subgoals or validation requires extra information then the tactic will have extra arguments.

The simplest tactical is `ORELSE`. It tries `tac1` and tests whether it fails by catching the corresponding exception (say `tactic`):

```
fun ORELSE (tac1,tac2) g =
    (tac1 g)  handle  tactic => (tac2 g);
```

The tactical `THEN` keeps track of many lists. It applies `tac1` to the goal, getting a list of subgoals $G_1, \ldots, G_n$ and validation $P$. It applies `tac2` to $G_1, \ldots, G_n$, getting a list of lists and a list of validations $P_1, \ldots, P_n$. Combining the validations involves tedious code. The final validation takes a list of theorems and partitions them into a list of lists $L_1, \ldots, L_n$, each of correct length for its corresponding validation. It applies each $P_i$ to $L_i$ and applies $P$ to the resulting theorems.

The following trivial tactic succeeds on all goals, mapping a goal to itself. It is an identify for `THEN`. The validation maps a one-element list to its sole element.

```
fun ALL_TAC g =  ([g],  fn [th] => th);
```

Tactical `REPEAT` is a recursive function that calls `ALL_TAC`.[10]

```
fun REPEAT tac g =
   ((tac THEN REPEAT tac) ORELSE ALL_TAC) g;
```

Validation-based tactics are flexible because subgoals and proofs are generated by arbitrary functions, but this flexibility has a cost. There is no way of detecting an invalid tactic until it is too late. When we attempt to use the validation it may fail, diverge, or return something unwanted. Each primitive tactic must be written individually.

The validation model above does not handle unification. Goals may not contain unknowns to be instantiated later. As a consequence, the LCF user must supply an explicit term at each $\exists$:right step. For $\forall$:left there are tactics that try to use universally quantified assumptions by matching. Sokołowski [1987] wrote a set of Edinburgh LCF tactics permitting unification. They maintain an environment of variable instantiations.

---

[10]`REPEAT` mentions goal `g` just to delay evaluation. Otherwise `REPEAT tac` would loop immediately, due to ML's strict semantics.

My recent book on LCF describes rules and tactics, as well as some theory and applications [Paulson 1987]. The Higher-Order Logic (HOL) prover, which is based on LCF, is coming into widespread use for hardware verification [Gordon 1988]. Several complicated chips have been verified using HOL. One of the largest HOL proofs concerns the Viper microprocessor [Cohn 1989a].

Nuprl, which is another relative of LCF, has unusual tactics that manipulate proof objects [Constable et al. 1986]. Nuprl implements a constructive logic based on Martin-Löf Type Theory, and has performed many proofs; in hardware verification, for example, [Basin and Del Vecchio 1990] have verified part of a floating point processor.

The next section describes the theorem prover Isabelle, which takes LCF's ideas in a different direction. Isabelle has been used in several laboratories for program verification and synthesis. Philippe Noël [1989] has formalized large pieces of Zermelo-Fraenkel set theory. Tobias Nipkow [1989], in a case study in data type refinement, has verified both sequential and distributed implementations of a memory system.

## 6.5 State-based tactics

Isabelle is a *generic* theorem prover: it supports reasoning in various logics [Paulson 1990]. It aims to minimize the amount of effort involved in adding a new logic. Neither rules nor their corresponding tactics require programming.

Isabelle represents rules as assertions, not as functions. The rule that takes $\phi_1, \ldots, \phi_n$ to $\phi$, written

$$[\phi_1, \ldots, \phi_n] \Rightarrow \phi$$

is a syntactic object. Its structure can be examined, which is impossible when rules are represented by functions. For instance, the conjunction rules might be represented as follows:

$$[?A, ?B] \Rightarrow ?A \wedge ?B \qquad [?A \wedge ?B] \Rightarrow ?A \qquad [?A \wedge ?B] \Rightarrow ?B$$

Meta-variables $?A$ and $?B$ can be replaced by formulae, producing instances of the rules.

What about rules with provisos, such as quantifier rules? An LCF rule can do any computation to check syntactic conditions. Isabelle has ways of handling provisos like '$x$ not free in $A$', but other provisos may become extra premises — and must be proved rather than checked.

A rule also represents the state of a backwards proof. Thus $[\phi_1, \ldots, \phi_n] \Rightarrow \phi$ represents the proof where the final goal is $\phi$ and the current subgoals are $\phi_1, \ldots, \phi_n$. The initial proof state is the trivial rule $[\phi] \Rightarrow \phi$. A typical backwards step replaces some $\phi_i$ by new subgoals $\psi_1, \ldots, \psi_m$. When $m = 0$, the total number of subgoals decreases. A successful proof terminates with simply $\phi$; this proof state is itself the desired theorem.

For example, the proof sketched in the previous section might produce the following sequence of states:
$$[G_1] \Rightarrow G_1$$
$$[G_2] \Rightarrow G_1$$
$$[G_{31}, G_{32}] \Rightarrow G_1 \quad (*)$$
$$[G_4, G_{32}] \Rightarrow G_1$$
$$[G_{32}] \Rightarrow G_1$$
$$G_1$$

Here we have always chosen the first subgoal. Choosing $G_{32}$ before $G_{31}$ at state $(*)$ would produce $[G_{31}] \Rightarrow G_1$ as the next state.

A backwards step using the rule $[\psi_1, \ldots, \psi_m] \Rightarrow \psi$ matches its conclusion, $\psi$, with the chosen subgoal, $\phi_i$. The new subgoals are the corresponding instances of the rule's premises $\psi_1, \ldots, \psi_m$. If $s$ is a substitution such that $\psi s = \phi_i$ then Isabelle permits the following meta-level inference:

$$\frac{[\psi_1, \ldots, \psi_m] \Rightarrow \psi \quad [\phi_1, \ldots, \phi_n] \Rightarrow \phi}{[\phi_1, \ldots, \phi_{i-1}, \psi_1 s, \ldots, \psi_m s, \phi_{i+1}, \ldots, \phi_n] \Rightarrow \phi}$$

Note that $\Rightarrow$ is essentially implication, so (object) rules are Horn clauses and the above inference is a form of resolution. It is a small step from matching to unification. If the proof state contains meta-variables then resolution involves unifying $\psi$ with $\phi_i$. If $\psi s = \phi_i s$ then resolution makes a new proof state by applying $s$ throughout:

$$[\phi_1 s, \ldots, \phi_{i-1} s, \psi_1 s, \ldots, \psi_m s, \phi_{i+1} s, \ldots, \phi_n s] \Rightarrow \phi s$$

Thus variables in the proof state can be instantiated, even if they are shared between several subgoals.

An Isabelle tactic is a function that maps one proof state to another. It takes not one goal but the entire proof state, and may change it in any way. Updating variables in the state can alter other subgoals and even the final goal. By starting with a goal that contains variables, we can use an Isabelle proof to compute answers. The proof state is a meta-theorem stating that the subgoals imply the goal: no validations are necessary.

Tacticals are implemented differently from LCF's. The tactic `tac1 THEN tac2` applies `tac1` to a proof state, and then applies `tac2` to the result. Tactic `ALL_TAC` returns an unchanged proof state. Tacticals `ORELSE` and `REPEAT` are like in LCF. Resolution provides all the primitive tactics for free.

## 6.6 Technical aspects of Isabelle

Some technical points about nondeterminism and quantifiers were omitted from the previous section for simplicity.

Tactical `ORELSE` does not permit backtracking. Consider the tactic

```
(tac1 ORELSE tac2)  THEN  tac3
```

If `tac1` succeeds then the `ORELSE` will choose it. If `tac3` then fails, the `ORELSE` cannot try `tac2`; instead, the entire tactic fails. To take account of the success of `tac3`, use the tactic

```
(tac1 THEN tac3)   ORELSE   (tac2 THEN tac3)
```

To solve this problem more generally, Isabelle tactics incorporate nondeterminism. A tactic returns a list of next states. An empty list means the tactic has failed. The tactic `tac1 THEN tac2` returns all states that could result by applying `tac1` followed by `tac2`. A new tactical, `APPEND`, is a nondeterministic `ORELSE`. Calling `tac1 APPEND tac2` returns all states that could result by applying `tac1` or `tac2`.

Since a tactic can return a list of outcomes, repetition of a tactic can produce a tree of states. Tactical `DEPTH_FIRST` performs a depth-first search of this tree, giving Prolog-style backtracking. Other strategies are available, including breadth-first and best-first (search guided by a distance function). The next state list must be lazy,[11] for it could be infinite, and only a few elements will be needed.

Complex techniques are required for quantifier rules:

$$[\forall x \,.\, ?F(x)] \Rightarrow ?F(?t)$$

Here $?F$ is a function variable. If it gets instantiated with the $\lambda$-abstraction $\lambda x.A$, then $?F(?t)$ causes the substitution of $?t$ into $A$. Ordinary unification cannot instantiate function variables or perform substitutions. Isabelle uses higher-order unification [Huet 1975], which is powerful but sometimes ill-behaved. Perhaps something simpler should be adopted; second-order matching is decidable [Huet and Lang 1978].

Originally Isabelle enforced quantifier provisos using parameters resembling Folderol's, and took resolution as a primitive meta-rule. Now Isabelle takes higher-order logic as its meta-logic [Paulson 1989]. This permits provisos to be enforced using meta-quantifiers and function variables, a dual of Skolemization. Derived forms of meta-reasoning include resolution and methods for natural deduction.

# 7 Conclusion

Folderol has achieved its purpose if it has illustrated the prospects, hazards, and techniques of writing a theorem prover. Although Folderol can prove some complex theorems, there is much that it cannot do. If you have a particular verification task in mind, you will probably see at once that Folderol is not up to the job.

A real theorem prover must include numbers, functions, pairs, and other mathematical concepts. It must provide flexible methods of proof discovery. It must maintain a database of symbols, axioms, theorems, and maybe proofs. We have surveyed other systems to see how they meet these requirements.

You may now recognize that writing a theorem prover is a mammoth task. My final advice is this. Don't write a theorem prover. Try to use someone else's.

---

[11]Standard ML uses strict evaluation, but lazy lists can be implemented using closures.

**Acknowledgement.** Thomas Forster, Mike Fourman, and Robin Milner commented on this chapter. LCF, ML, and Isabelle were all developed by funding from the Science and Engineering Research Council (grants GR/B67766, GRT/D 30235, and GR/E0355.7, among others).

# 8 Program listing

```
(*BASIC FUNCTIONS*)

(*Length of a list*)
local fun length1 (n, [ ])  = n
        | length1 (n, x::l) = length1 (n+1, l)
in  fun length l = length1 (0,l) end;

(*The elements of a list satisfying the predicate p.*)
fun filter p [] = []
  | filter p (x::xs) = if p(x) then  x :: filter p xs  else  filter p xs;

infix mem;  (*membership in a list*)
fun x mem []  =  false
  | x mem (y::l)  =  (x=y) orelse (x mem l);

infix ins;  (*insertion into list if not already there*)
fun x ins xs = if x mem xs then   xs    else   x::xs;

fun repeat x 0 = []
  | repeat x n = x :: (repeat x (n-1));

fun accumulate f ([], y) = y
  | accumulate f (x::xs, y) = accumulate f (xs, f(x,y));

(*Look for a pair (X,z) in environment, return [z] if found, else [] *)
fun lookup (X, []) = []
  | lookup (X, (Y,z)::env) = if X = (Y:string) then [z]  else  lookup(X,env);

exception ERROR of string;

(*TERMS AND FORMULAE*)

datatype term =
    Var   of string
  | Param of string * string list
  | Bound of int
  | Fun   of string * term list;

datatype form =
    Pred  of string * term list
  | Conn  of string * form list
  | Quant of string * string * form;

(*variables
```

```
       a,b,c: string      q,r: string (quantifier names)
       i,j: int     (Bound indexes)
       t,u: term    A,B: form
       x,y: any     f,g: functions
*)

(*Operations on terms and formulae*)

(*Replace the atomic term u by new throughout a term*)
fun replace_term (u,new) t =
    if t=u then new else
    case t of Fun(a,ts) => Fun(a, map (replace_term(u,new)) ts)
            | _ => t;

(*Abstraction of a formula over u (containing no bound vars).*)
fun abstract t =
    let fun abs i (Pred(a,ts)) = Pred(a, map (replace_term (t, Bound i)) ts)
          | abs i (Conn(b,As)) = Conn(b, map (abs i) As)
          | abs i (Quant(q,b,A)) = Quant(q, b, abs (i+1) A)
    in  abs 0  end;

(*Replace (Bound 0) in formula with t (containing no bound vars).*)
fun subst_bound t =
    let fun subst i (Pred(a,ts)) = Pred(a, map (replace_term (Bound i, t)) ts)
          | subst i (Conn(b,As)) = Conn(b, map (subst i) As)
          | subst i (Quant(q,b,A)) = Quant(q, b, subst (i+1) A)
    in  subst 0  end;

(*SYNTAX: SCANNING, PARSING, AND DISPLAY*)

(*Scanning a list of characters into a list of tokens*)

datatype token = Key of string  |  Id of string;

fun is_char(l,c,u) = ord l <= ord c  andalso  ord c <= ord u;

fun is_letter_or_digit c =
    is_char("A",c,"Z") orelse is_char("a",c,"z") orelse is_char("0",c,"9");

(*Scanning of identifiers and keywords*)

fun token_of a = if a mem ["ALL","EXISTS"]  then  Key(a)  else  Id(a);

fun scan_ident (front, c::cs) =
        if is_letter_or_digit c
        then  scan_ident (c::front, cs)
        else  (token_of (implode(rev front)), c::cs)
  | scan_ident (front, []) = (token_of (implode(rev front)), []);

(*Scanning, recognizing --> and <->, skipping blanks, etc.*)
fun scan (front_toks, []) = rev front_toks     (*end of char list*)
```

```
            (*long infix operators*)
  | scan (front_toks, "-"::"-"::">"::cs) = scan (Key"-->" ::front_toks,  cs)
  | scan (front_toks, "<"::"-"::">"::cs) = scan (Key"<->" ::front_toks,  cs)
            (*blanks, tabs, newlines*)
  | scan (front_toks, " "::cs) = scan (front_toks,  cs)
  | scan (front_toks, "\t"::cs) = scan (front_toks,  cs)
  | scan (front_toks, "\n"::cs) = scan (front_toks,  cs)
  | scan (front_toks, c::cs) =
      if is_letter_or_digit c then scannext(front_toks, scan_ident([c], cs))
      else  scan (Key(c)::front_toks,  cs)
and scannext (front_toks, (tok, cs)) = scan (tok::front_toks,  cs);

(*Parsing a list of tokens*)

fun apfst f (x,toks) = (f x, toks);

(*Functions for constructing results*)
fun cons x xs = x::xs;
fun makeFun fu ts = Fun(fu,ts);
fun makePred id ts = Pred(id,ts);
fun makeNeg A = Conn("~", [A]);
fun makeConn a A B = Conn(a, [A,B]);
fun makeQuant q b A = Quant(q, b, abstract (Fun(b,[])) A);

(*Repeated parsing, returning the list of results  *)
fun parse_repeat (a,parsefn) (Key(b)::toks) = (*    a<phrase>...a<phrase>   *)
          if a=b then parse_repeat1 (a,parsefn) toks
          else ([], Key(b)::toks)
  | parse_repeat (a, parsefn) toks = ([], toks)
and parse_repeat1 (a,parsefn) toks =               (*    <phrase>a...a<phrase>   *)
      let val (u,toks2) = parsefn toks
      in   apfst (cons u) (parse_repeat (a, parsefn) toks2)  end;

fun rightparen (x, Key")"::toks) = (x, toks)
  | rightparen _ = raise ERROR "Symbol ) expected";

fun parse_term (Id(a)::Key"("::toks) =
        apfst (makeFun a) (rightparen (parse_repeat1 (",", parse_term) toks))
  | parse_term (Id(a)::toks) = (Fun(a,[]), toks)
  | parse_term (Key"?"::Id(a)::toks) = (Var a, toks)
  | parse_term _ = raise ERROR "Syntax of term";

(*Precedence table*)
fun prec_of "~"   = 4
  | prec_of "&"   = 3
  | prec_of "|"   = 2
  | prec_of "<->" = 1
  | prec_of "-->" = 1
  | prec_of _     = ~1    (*means not an infix*);
```

```
(*Parsing of formulae;   prec is the precedence of the operator to the left;
    parsing stops at an operator with lower precedence*)
fun parse (Key"ALL"   ::Id(a)::Key".":: toks) =
      apfst (makeQuant "ALL" a) (parse toks)
  | parse (Key"EXISTS"::Id(a)::Key".":: toks) =
      apfst (makeQuant "EXISTS" a) (parse toks)
  | parse toks = parsefix 0 (parse_atom toks)
and parsefix prec (A, Key(co)::toks) =
      if prec_of co < prec then (A, Key(co)::toks)
      else parsefix prec
              (apfst (makeConn co A)
                  (parsefix (prec_of co) (parse_atom toks)))
  | parsefix prec (A, toks) = (A, toks)
and parse_atom (Key"~"::toks) = apfst makeNeg (parse_atom toks)
  | parse_atom (Key"("::toks) = rightparen (parse toks)
  | parse_atom (Id(pr)::Key"("::toks) =
        apfst (makePred pr) (rightparen (parse_repeat1 (",", parse_term) toks))
  | parse_atom (Id(pr)::toks) = (Pred(pr,[]), toks)
  | parse_atom _ = raise ERROR "Syntax of formula";

(*check that no tokens remain*)
fun parse_end (x, []) = x
  | parse_end (_, _::_) = raise ERROR "Extra characters in formula";

fun read a = parse_end (parse (scan([], explode a)));

(*Printing: conversion of terms/formulae to strings*)

fun enclose a = "(" ^ a ^ ")";

fun conc_list sep [] = ""
  | conc_list sep (b::bs) = (sep ^ b) ^ (conc_list sep bs);

fun conc_list1 sep (b::bs) = b ^ (conc_list sep bs);

fun stringof_term (Param(a,_)) = a
  | stringof_term (Var a) = "?"^a
  | stringof_term (Bound i) = "B." ^ makestring i
  | stringof_term (Fun (a,ts)) =  a ^ stringof_args ts
and stringof_args [] = ""
  | stringof_args ts = enclose (conc_list1 "," (map stringof_term ts));

fun max(m,n) : int = if m>n then m else n;

fun stringof_form prec (Pred (a,ts)) =  a ^ stringof_args ts
  | stringof_form prec (Conn("~", [A])) =   "~" ^ stringof_form (prec_of "~") A
```

```
  | stringof_form prec (Conn(C, [A,B])) =
        let val stringf = stringof_form (max(prec_of C, prec));
            val Z = stringf A ^ " " ^ C ^ " " ^ stringf B
        in  if (prec_of C <= prec)  then  (enclose Z)  else Z
        end
  | stringof_form prec (Quant(q,b,A)) =
        let val B = subst_bound (Fun(b,[])) A
            val Z = q^" "^b ^ ". " ^ stringof_form 0 B
        in  if  prec>0  then  (enclose Z)  else Z
        end
  | stringof_form prec _ = raise ERROR "stringof_form: Bad formula";

val stringof = stringof_form 0;

(*UNIFICATION*)

exception UNIFY;

(*Naive unification of terms containing no bound variables*)
fun unify_terms ([],[], env) = env
  | unify_terms (t::ts, u::us, env) =
      let fun chasevar (Var a) =  (*Chase variable assignments*)
                (case  lookup(a,env)  of
                    u::_ => chasevar u  |  [] => Var a)
            | chasevar t = t;
          fun unify_var (a, t) = (*unification with var*)
              let fun occs (Fun(_,ts)) = occsl ts
                    | occs (Param(_,bs)) = occsl(map Var bs)
                    | occs (Var b) =  a=b  orelse occsl(lookup(b,env))
                    | occs _ = false
                  and occsl [] = false
                    | occsl(t::ts) = occs t  orelse  occsl ts
              in  if t = Var a  then  env
                    else if occs t then  raise UNIFY  else  (a,t)::env
              end
          and unify_term (Var a, t) = unify_var (a, t)
            | unify_term (t, Var a) = unify_var (a, t)
            | unify_term (Param(a,_), Param(b,_)) =
                  if a=b then env  else  raise UNIFY
            | unify_term (Fun(a,ts), Fun(b,us)) =
                  if a=b then unify_terms (ts,us,env)  else  raise UNIFY
            | unify_term _ =  raise UNIFY
    in  unify_terms (ts, us, unify_term (chasevar t, chasevar u))  end
  | unify_terms _ =  raise UNIFY;

(*Unification of atomic formulae*)
fun unify (Pred(a,ts), Pred(b,us), env) =
        if a=b then unify_terms(ts,us,env)  else  raise UNIFY
  | unify _ =  raise UNIFY;
```

```
(*Accumulate all Vars in the term (not Vars attached to a Param).*)
fun vars_in_term (Var a, bs) = a ins bs
  | vars_in_term (Fun(_,ts), bs) = accumulate vars_in_term (ts,bs)
  | vars_in_term (_, bs) = bs;

(*Instantiate a term by an environment*)
fun inst_term env (Fun(a,ts)) = Fun(a, map (inst_term env) ts)
  | inst_term env (Param(a,bs)) =
        Param(a, accumulate vars_in_term (map (inst_term env o Var) bs, []))
  | inst_term env (Var a) =
      (case  lookup(a,env)  of
           u::_ =>  inst_term env u
         | []   =>  Var a)
  | inst_term env t = t;

(*INFERENCE: GOALS AND PROOF STATES: GOALS AND PROOF STATES*)

datatype side = Left | Right;

type entry = int * side * form;
type goal = entry list;
type goaltable = goal list;

fun inst_form [] A = A
  | inst_form env (Pred(a,ts))   = Pred(a, map (inst_term env) ts)
  | inst_form env (Conn(b,As))   = Conn(b, map (inst_form env) As)
  | inst_form env (Quant(q,b,A)) = Quant(q, b, inst_form env A);

fun inst_goal env [] = []
  | inst_goal env ((m,si,A)::G) = (m, si, inst_form env A) :: inst_goal env G;

fun inst_goals [] Gs = Gs
  | inst_goals env Gs = map (inst_goal env) Gs : goaltable;

(*Accumulate over all terms in a formula*)
fun accum_form f (Pred(_,ts), bs) = accumulate f (ts, bs)
  | accum_form f (Conn(_,As), bs) = accumulate (accum_form f) (As, bs)
  | accum_form f (Quant(_,_,A), bs) = accum_form f (A,bs);

(*Accumulate over all formulae in a goal*)
fun accum_goal f ([], bs) = bs
  | accum_goal f ((_,_,A)::G, bs) = accum_goal f (G, f(A,bs));

val vars_in_form = accum_form vars_in_term;
val vars_in_goal = accum_goal vars_in_form;

(*Accumulate all Params*)
fun params_in_term (Param (a,bs), pairs) = (a,bs) ins pairs
  | params_in_term (Fun(_,ts), pairs) = accumulate params_in_term (ts, pairs)
```

```
      | params_in_term (_, pairs) = pairs;

val params_in_form = accum_form params_in_term;
val params_in_goal = accum_goal params_in_form;

(*Returns (As,Bs),preserving order of elements
  As = Left entries,  Bs = Right entries *)
fun split_goal G =
    let fun split (As,Bs, []: goal) = (As,Bs)
          | split (As,Bs, (_,Left,A)::H) = split (A::As,Bs, H)
          | split (As,Bs, (_,Right,B)::H) = split (As, B::Bs, H)
    in  split([], [], rev G)  end;

fun is_pred (Pred _) = true
  | is_pred _ = false;

(*Solve the goal (A|-A') by unifying A with A', Left and Right atomic formulae.
  Returns list [ (A,env) ] if successful, otherwise []. *)
fun solve_goal G =
    let fun findA ([], _) = []      (*failure*)
          | findA (A::As, Bs) =
              let fun findB [] = findA (As,Bs)
                    | findB (B::Bs) = [ (A, unify(A,B,[])) ]
                          handle UNIFY => findB Bs
              in  findB Bs  end
        val (As,Bs) = split_goal G
    in  findA(filter is_pred As, filter is_pred Bs)  end;

(*Insert goals into a goaltable.  For each solved goal (A,env),
  accumulates the formula (in reverse) and instantiates all other goals.*)
fun insert_goals ([], As, tab) = (As,tab)
  | insert_goals (G::Gs, As, tab) =
      case  solve_goal G  of
          (A,env)::_ =>         (*instantiate other goals*)
            insert_goals (inst_goals env Gs,
                          (inst_form env A) :: As,
                          inst_goals env tab)
        | [] =>  insert_goals (Gs, As, G::tab);

fun stringof_sy (Pred(a,_)) = a
  | stringof_sy (Conn(a,_)) = a
  | stringof_sy (Quant(q,_,_)) = q;

fun stringof_side Right = ":right"
  | stringof_side Left = ":left";

(*Generation of new variable names*)
local
  fun make_letter n = chr(ord("a")+n);
```

```
    fun make_varname (n,tail) =
        if n<26 then make_letter n ^ tail
        else make_varname (n div 26, make_letter(n mod 26) ^ tail);
    val varcount = ref ~1
in
fun gensym() = (varcount := !varcount+1;  make_varname (!varcount,""))
and init_gensym() = varcount := ~1
end;

(*The "cost" of reducing a connective*)
fun cost (_,     Conn("~", _))        = 1       (*a single subgoal*)
  | cost (Left,  Conn("&", _))        = 1
  | cost (Right, Conn("|", _))        = 1
  | cost (Right, Conn("-->", _))      = 1
  | cost (Right, Quant("ALL",_,_))    = 1
  | cost (Left,  Quant("EXISTS",_,_)) = 1
  | cost (Right, Conn("&", _))        = 2       (*case split: 2 subgoals*)
  | cost (Left,  Conn("|", _))        = 2
  | cost (Left,  Conn("-->", _))      = 2
  | cost (_    , Conn("<->", _))      = 2
  | cost (Left,  Quant("ALL",_,_))    = 3       (*quantifier expansion*)
  | cost (Right, Quant("EXISTS",_,_)) = 3       (*quantifier expansion*)
  | cost _ = 4 ;                                (*no reductions possible*)

fun paircost (si,A) = (cost(si,A), si, A);

(*Insertion into a list, ordered by sort keys. *)
fun insert less =
  let fun insr (x, []) = [x]
        | insr (x, y::ys) = if less(y,x) then y :: insr (x,ys) else x::y::ys
  in  insr  end;

(*Insert an entry into a goal, in correct order *)
fun entry_less ((m,_,_): entry, (n,_,_): entry) = m<n;
val insert_early = insert entry_less;

(*Quantified formulae are put back at end -- they are used in a cycle*)
fun entry_lesseq ((m,_,_): entry, (n,_,_): entry) = m<=n;
val insert_late  = insert entry_lesseq;

(*Extend the goal G by inserting a list of (side,form) pairs*)
fun new_goal G pairs = accumulate insert_early (map paircost pairs, G);

(*Extend the goal G, making a list of goals*)
fun new_goals G pairslist = map (new_goal G) pairslist;

exception REDUCE;

(*Reduce the pair using the rest of the goal (G) to make new goals*)
fun reduce_goal (pair, G) =
  let val goals = new_goals G;
```

```
          fun vars_in A = vars_in_goal (G, vars_in_form(A,[]));
          fun subparam A = subst_bound (Param(gensym(), vars_in A)) A;
          fun subvar A   = subst_bound (Var(gensym())) A;
          fun red(_,Right,Conn("~", [A]))   = goals[[(Left,A)]]
            | red(_,Left, Conn("~", [A]))   = goals[[(Right,A)]]
            | red(_,Right,Conn("&", [A,B])) = goals[[(Right,A)], [(Right,B)]]
            | red(_,Left, Conn("&", [A,B])) = goals[[(Left,A),(Left,B)]]
            | red(_,Right,Conn("|", [A,B])) = goals[[(Right,A),(Right,B)]]
            | red(_,Left, Conn("|", [A,B])) = goals[[(Left,A)], [(Left,B)]]
            | red(_,Right,Conn("-->", [A,B])) = goals[[(Left,A),(Right,B)]]
            | red(_,Left, Conn("-->", [A,B])) = goals[[(Right,A)], [(Left,B)]]
            | red(_,Right,Conn("<->", [A,B])) =
                   goals[[(Left,A),(Right,B)], [(Right,A),(Left,B)]]
            | red(_,Left, Conn("<->", [A,B])) =
                   goals[[(Left,A),(Left,B)], [(Right,A),(Right,B)]]
            | red(_,Right,Quant("ALL",_,A)) = goals[[(Right, subparam A)]]
            | red(_,Left, Quant("ALL",_,A)) =
                  [ insert_early (paircost(Left, subvar A), insert_late(pair,G)) ]
            | red(_,Right,Quant("EXISTS",_,A)) =
                  [ insert_early (paircost(Right, subvar A), insert_late(pair,G)) ]
            | red(_,Left, Quant("EXISTS",_,A)) = goals[[(Left, subparam A)]]
            | red _ = raise REDUCE
      in  red pair  end;

(*Print the string a*)
fun prints a = output(std_out,a);

(*Print the rule used, with each formula found by unification,
     indenting by number of goals left.*)
fun print_step ((_,si,C), ngoals, As) =
   (prints (implode(repeat " " ngoals) ^
                 stringof_sy C ^ stringof_side si);
    prints (conc_list "   " (map stringof (rev As)));  prints"\n");

(*A single inference in the goaltable*)
fun proof_step [] = [] : goaltable
  | proof_step ([]::tab) = raise ERROR "Empty goal"
  | proof_step ((ent::G)::tab) =
       let val (As,newtab) = insert_goals (reduce_goal(ent,G), [], tab)
       in  print_step(ent, length tab, As);   newtab   end;

(*Perform n proof steps, no limit if n<0.  Stops if impossible to continue.*)
fun proof_steps (_,[]) = []    (*success -- no goals*)
  | proof_steps (0,tab) = tab
  | proof_steps (n,tab) = proof_steps (n-1, proof_step tab)
      handle REDUCE => (prints"\n**No proof rules applicable**\n";  tab);

fun pair si A = (si,A);

(*Make a goal from lists of formulae: As|-Bs*)
fun make_goal (As,Bs) : goal =
    new_goal [] (map (pair Left) As  @   map (pair Right) Bs);
```

```
(*Reading of goals: Astrs|-Bstrs *)
fun read_tab (Astrs,Bstrs) : goaltable =
    let val As = rev(map read Astrs)
        and Bs = rev(map read Bstrs);
        val G = make_goal(As,Bs);
        val (_, tab) = insert_goals ([G],  [],  [])
    in  tab  end;

fun stringof_sequent [] = "empty"
  | stringof_sequent As = conc_list1 ", " (map stringof As);

fun print_goal G =
    let val (As,Bs) = split_goal G
    in  prints (stringof_sequent As ^ "  |-  " ^ stringof_sequent Bs ^ "\n\n")
    end;

fun print_param (a,ts) =
      prints (a ^ "          " ^ stringof_args (map Var ts) ^ "\n");

fun print_params [] = ()
  | print_params pairs =
      (prints "Param     Not allowed in\n";
       map print_param pairs;  prints "\n");

fun print_count 1 = ()
  | print_count n = prints (makestring n ^ " goals\n");

fun print_tab [] = prints"No more goals: proof finished\n"
  | print_tab Gs =
     (prints"\n";   map print_goal Gs;  print_count (length Gs);
      print_params (accumulate params_in_goal (Gs,[])));

(*Top-level commands: interaction with proof state*)

val the_goaltable = ref ([] : goaltable);

fun set_tab tab = (the_goaltable := tab;  print_tab tab);

(*Read a goal: the sequent As|-Bs *)
fun read_goalseq (Astrs,Bstrs) =
    (init_gensym();   set_tab(read_tab (Astrs,Bstrs)));

(*Read the goal |-B *)
fun read_goal Bstr = read_goalseq([],[Bstr]);

fun step()   = set_tab (proof_step(!the_goaltable));
fun steps n  = set_tab (proof_steps (max(n,0), !the_goaltable));
fun run()    = set_tab (proof_steps (~1, !the_goaltable));

fun run_goalseq (Astrs,Bstrs) = (read_goalseq (Astrs,Bstrs);  run());
fun run_goal b = run_goalseq([],[b]);
```

```
(*Raises exception unless some goals are left unsolved after n proof steps*)
fun fail_goal n A =
   (read_goal A;
    steps n;
    (case !the_goaltable of
         [] => raise ERROR "This proof should have failed!"
       | _::_ => prints"Failed, as expected\n"));
```